\documentclass[11pt,fleqn]{article}

\usepackage{graphicx}
\usepackage{a4wide}
\usepackage{epsfig}
\usepackage{color}
\usepackage{latexsym}
\usepackage{amssymb}
\usepackage{amsmath}
\usepackage{xspace}
\usepackage{url}
\usepackage{times}

\DeclareSymbolFont{msbm}{U}{msb}{m}{n}
\DeclareMathSymbol{\C}{\mathalpha}{msbm}{'103}
\DeclareMathSymbol{\R}{\mathalpha}{msbm}{'122}
\DeclareMathSymbol{\Q}{\mathalpha}{msbm}{'121}
\DeclareMathSymbol{\Z}{\mathalpha}{msbm}{'132}
\DeclareMathSymbol{\N}{\mathalpha}{msbm}{'116}
\DeclareMathSymbol{\K}{\mathalpha}{msbm}{'113}

\newtheorem{lemma}{Lemma}
\newtheorem{cor}[lemma]{Corollary}
\newtheorem{theorem}[lemma]{Theorem}

\newcommand{\proof}{\noindent\textbf{Proof.}\enspace}

\newcommand{\old}[1]{{}}

\newcommand{\qed}{\hfill{\ensuremath{\Box}}}
\newcommand{\eps}{\varepsilon}

\newcommand{\NP}{\ensuremath{\mathcal{N}\mathcal{P}}\xspace}

\newcommand{\bi}{\begin{itemize}}
\newcommand{\ei}{\end  {itemize}}
\newcommand{\bt}{\begin{tabbing}}
\newcommand{\et}{\end  {tabbing}}
\newcommand{\be}{\begin{enumerate}}
\newcommand{\ee}{\end  {enumerate}}

\newcommand{\all}{} 
\newcommand{\Ms}[2][P]{\ensuremath{\mbox{\it St-Mat}_{#2}(#1)}}
\newcommand{\Ts}[2][P]{\ensuremath{\mbox{\it St-Tre}_{#2}(#1)}}
\newcommand{\TTs}[2][P]{\ensuremath{\mbox{\it St-{\ensuremath\Delta}}_{#2}(#1)}}

\newcommand{\CMs}[2][P]{\ensuremath{\mbox{\it Cr-Mat}_{#2}(#1)}}
\newcommand{\CTs}[2][P]{\ensuremath{\mbox{\it Cr-Tre}_{#2}(#1)}}
\newcommand{\CTTs}[2][P]{\ensuremath{\mbox{\it Cr-{\ensuremath\Delta}}_{#2}(#1)}}

\newcommand{\eg}{e.g., }

\title{Minimizing the Stabbing Number of Matchings, Trees, and
  Triangulations\thanks{An extended abstract
appeared in the {\em Proceedings of the 15th ACM-SIAM Symposium on Discrete Algorithms \cite{flm-msnmstt-04}.}}}  
\author{
S\'andor P.\ Fekete\thanks{
      Algorithms Group, Department of Computer Science, Braunschweig
      University of Technology,
      M\"uhlenpfordtstr.\ 23,  D-38106 Braunschweig, Germany.
      Email: \texttt{s.fekete@tu-bs.de}.
      }
\and
Marco E.\ L\"ubbecke\thanks{
      Institut f\"ur Mathematik,
      Sekr.\ MA 5-1,
      Technische Universit\"at Berlin,
      Stra{\ss}e des 17.\ Juni 136,  D-10623 Berlin, Germany.
      Email: \texttt{m.luebbecke@math.tu-berlin.de}.
      Visits to Kingston and Stony Brook were supported by a DFG
      travel grant.
      }
\and
Henk Meijer\thanks{
      Department of Science,
      Roosevelt Academy,
      Middelburg (ZL), The Netherlands. 
      Email: \texttt{h.meijer@roac.nl}.
      Partially supported by NSERC while visiting Braunschweig in 2002.
      }
}
\date{}

\begin{document}
\maketitle

\begin{abstract}
  The (axis-parallel) stabbing number of a given set of line segments
  is the maximum number of segments that can be intersected by any one
  (axis-parallel) line. This paper deals with finding perfect
  matchings, spanning trees, or triangulations of minimum stabbing
  number for a given set of vertices.  The complexity of finding a 
  spanning tree of minimum stabbing number 
  is one of the original 30 questions on ``The Open Problems Project''
  list of outstanding problems in computational 
  geometry by Demaine, Mitchell, and O'Rourke.

  We show \NP-hardness of stabbing problems 
  by means of a general proof technique. 
  For matchings, this also implies a non-trivial lower bound on the
  approximability. On the positive side, we propose a cut based integer 
  programming formulation for
  minimizing the stabbing number of matchings and spanning trees. From the
  corresponding linear programming relaxation we
  obtain polynomial-time lower bounds and show that there always
  is an optimal fractional solution that contains an edge of at least 
  constant weight. We conjecture that the resulting 
  iterated rounding scheme constitutes a constant-factor approximation
  algorithm.
  
\end{abstract}


{\bf ACM Classification:} F.2.2 Nonnumerical Algorithms and Problems.

{\bf AMS Classification:} 68Q17, 68U05, 90C27.

{\bf Keywords:} Stabbing number, ma\-tching, spanning tree,
triangulation, complexity, linear programming relaxation, iterated rounding.

\section{Introduction}
\label{sec:intro}
\paragraph{Objective Functions.}
Many problems in combinatorial optimization, algorithmic graph
theory, or computational geometry deal with minimizing the length 
of a desired structure: given a set of vertices, find a set
of line segments of small total length, such that a certain structural
condition is maintained.  Among the most popular structures
are spanning trees, perfect matchings, or (in a
planar geometric setting) triangulations of minimum total length. 
Other geometric problems give rise to other objective functions:
for example, one can ask for the total turn cost between
adjacent line segments; \eg see
\cite{abdfms-octtc-05}.

When dealing with structural or algorithmic properties, 
another possible objective function is the {\em
  stabbing number}: 
for a given set of line segments, this is the
maximum number of segments that are encountered (in their interior or
at an endpoint) by any line.
If we consider only axis-parallel lines, we get the {\em axis-parallel
  stabbing number}. 
A closely related measure defined by Matou{\v s}ek \cite{cg:Matousek:91}
is the {\em crossing number}, 
which is the number of connected components of
the intersection of a line with the union of line 
segments\footnote{This should not be confused with the crossing number
in graph drawing, which is the total number of crossing line segments.}.
If there are no connected components of
collinear segments (which is the case for matchings), the crossing
number coincides with the stabbing number. When considering structures
like triangulations, the crossing number is precisely one more than
the maximum number of triangles intersected by any one line.

Stabbing problems have been considered for several years. The
complexity of many algorithms in computational geometry is directly
dependent on the complexity of ray shooting; as described by Agarwal
\cite{cg:Agarwal:92}, the latter can be improved by making use of 
spanning trees of low stabbing number.
We will sketch some related results further down.
Most previous work on 
stabbing and crossing problems has focused on extremal
properties, and little has been known about the computational complexity
of actually finding structures of low stabbing number, or possible 
approximation algorithms. In fact, 
settling the complexity of Minimum Stabbing Number
for spanning trees has been one of the original 30 outstanding open
problems of computational geometry on the list by Mitchell and
O'Rourke~\cite{cg:MitchellORourke:01}. (An up-to-date list is
maintained online by Demaine, Mitchell, and
O'Rourke~\cite{cg:DemaineORourke:03}.) 

\paragraph{Our Contributions.}
We describe a general proof technique that shows
\NP-hardness of minimizing the stabbing
number of perfect matchings, triangulations, and spanning trees.  For the 
case of matchings we show that it is also hard to approximate 
the minimum stabbing number within a factor below 6/5. 

On the other hand, we present a mathematical programming
framework for actually finding structures with small stabbing number. 
Our approach characterizes solutions to stabbing problems
as integer programs (IPs) with an exponential
number of cut constraints. We describe how the corresponding linear
programming (LP) relaxations can be solved
in polynomial time, providing empirically excellent lower bounds.
Furthermore, we show that there always is an optimal
fractional matching (or spanning tree) that contains an edge of
weight above a lower bound of 1/3 (or 1/5 for spanning trees), allowing 
an iterated rounding scheme 
similar to the one developed by Jain for the
generalized Steiner network problem~\cite{jain}:
compute a heuristic solution by solving a polynomial
number of LPs. We conjecture that the objective function value of this
heuristic solution is within a constant factor
of the optimum. Our mathematical
programming approach is also practically useful:
as described in detail in our experimental study \cite{flm-csosn-07},
we can optimally solve stabbing problems for 
instances (taken from well-known benchmark sets of other
geometric optimization problems) of vertex sets up to several hundred vertices.

Our results in detail:
\begin{itemize}
\item We prove that deciding whether a vertex set has a perfect matching of 
  axis-parallel stabbing number 5 is an \NP-complete problem; 
  we also extend this result to general stabbing number.
\item We prove that finding a triangulation of minimum axis-parallel
  stabbing number is an \NP-hard problem; we also extend this
  result to general stabbing number.
\item We prove that finding a spanning tree of minimum
  axis-parallel stabbing number is an \NP-hard problem; we extend this
  result to general stabbing number, and sketch \NP-hardness
  proofs for minimum axis-parallel or general crossing number.
\item We give an IP-based formulation for stabbing problems; the
  corresponding fractional LP solutions can be
  computed in polynomial time, providing a family of lower bounds. 
\item We give results on the structure of fractional vertices of the resulting
  LP relaxation: for matching, we show that there always is
  an edge with weight at least $1/5$, while for spanning trees,
  there always is an edge with weight greater than $1/3$. This gives
  way to a heuristic algorithm based on iterated rounding;
  we conjecture that the resulting 
  solution values are within a constant factor of the optimum.
\end{itemize}

The vertex sets constructed in our hardness proofs
make critical use of the collinearity of vertices.
On the other hand, our positive (LP-based) results do {\em not} make any
assumptions on the structure of the vertex set: they can be used
for vertex sets in degenerate as well as in general position,
and can be applied to any family of stabbing lines that can be evaluated
by considering a subset of polynomially many representatives.

We have also performed a computational study on a diverse set of 
instances; the results show that our LP-based approach  is
good not only in theory (where we get a polynomial running time based on
the ellipsoid method), but also for actually solving instances in
practice (where we use the simplex method). Details are omitted
from this theoretical paper; a report on the practical results
can be found in \cite{flm-csosn-07}.

\paragraph{Related Work.}
Existing work dealing with structures of low stabbing number can be
divided into algorithmic applications and implications on one hand, and
extremal properties on the other hand.

Agarwal~\cite{cg:Agarwal:92} describes improved algorithmic
solutions for problems such as
ray shooting and implicit point locations queries (which by themselves
have applications in polygon containment, implicit hidden surface
removal, polygon placement, etc.); his main tool are
spanning trees with low stabbing number.
One of the theoretically best
performing data structures for ray tracing in two dimensions is based
on a triangulation of the polygonal scene; see Hershberger and
Suri~\cite{cg:HershbergerSuri:95}: in their ``pedestrian'' approach
to ray shooting, the complexity of a query is simply the number
of triangles visited, i.e., corresponds precisely to the stabbing number.
Held, Klosowski, and
Mitchell~\cite{cg:HeldKlosowskiMitchell:95} investigate collision
detection in a virtual reality environment, again, based on ``pedestrian''
ray shooting. More recently, Aronov et
al.~\cite{abcc-cdocs-05} have performed an experimental study of the
complexity of ray tracing algorithms and run-time predictors, which include
average number of intersection points for a transversal line, and
depth complexity.

Extremal properties of crossing numbers were considered by Welzl
\cite{cg:Welzl:92} and by Matou{\v s}ek~\cite{cg:Matousek:91}, 
who showed that any planar set of $n$ vertices has a
spanning tree with a crossing number of $O(\sqrt{n})$, and provided
examples requiring a crossing number of $\Omega(\sqrt{n})$.  Another
variant is studied by de Berg and van Kreveld
\cite{cg:deBergvanKreveld:94}: the stabbing number of a decomposition
of a rectilinear polygon $P$ into rectangles is the maximum number of
rectangles intersected by any axis-parallel segment that lies
completely inside of $P$; they prove that any simple rectilinear polygon
with $n$ vertices admits a decomposition with stabbing number $O(\log
n)$, and they give an example of a simple rectilinear polygon for
which any decomposition has stabbing number $\Omega(\log n)$. They
generalize their results to rectilinear polygons with rectilinear
holes.
Agarwal, Aronov, and Suri
\cite{cg:AgarwalAronovSuri:95}
investigate extremal properties of the stabbing number of
triangulations in three dimensions, where the stabbed objects are
simplices; see also  Aronov and Fortune~\cite{af-amwttd-99} for
this problem.
Shewchuk \cite{cg:Shewchuk:04} shows that in $d$ dimensions, a
line can stab the interiors of $\Theta(n^{\lceil d/2\rceil})$ Delaunay
$d$-simplices. This implies, in particular, that a Delaunay
triangulation in the plane may have linear stabbing number.
More recently, T\'oth~\cite{t-oslsn-05} showed that for any subdivision
of $d$-dimensional Euclidean space, $d\geq 2$, by $n$ axis-aligned
boxes, there is an axis-parallel line that stabs at least 
$\Omega(\log^{1/(d-1)}n)$ boxes, which is the best possible lower bound.
Generalizations of the stabbing objects have also been considered:
most notably, Chazelle and Welzl~\cite{cw-qorss-89} describe extremal
properties of stabbing spanning trees in $d$-dimensional space by hyperplanes;
the analogous problem and a corresponding result for matchings is 
also discussed in \cite{cw-qorss-89}.

\paragraph{This Paper.}
The rest of this paper is organized as follows.  After some basic
definitions and notation in Section~\ref{sec:prelim}, we give details
of our various hardness proofs in Section~\ref{sec:complex}. 
In Section~\ref{sec:LP}, we describe our LP-based approach for
constructing bounds. Section~\ref{sec:iterated} presents an iterated
rounding technique for matching and spanning tree problems; we believe that
the resulting algorithms yield constant-factor approximations. 
Final concluding thoughts and miscellaneous results and 
problems are presented in Section~\ref{sec:conc}.

\section{Preliminaries}
\label{sec:prelim}

Given a set $L$ of line segments in the plane, the \emph{stabbing
  number} of a line $\ell$ is the number of segments of $L$ that are
intersected by $\ell$. The \emph{stabbing number of} $L$ is the
maximum stabbing number over all lines $\ell$; the \emph{axis-parallel
  stabbing number of} $L$ is the maximum stabbing number over all
axis-parallel lines $\ell$.  In this paper, the set $L$ will arise as
the edges of a perfect matching, spanning tree, or triangulation of 
a given set $P$
of $n$ vertices in the plane, and our objective is to find such a
structure of minimum stabbing number. Any reference to matching always
means perfect matching.  Therefore, when dealing with matchings, we
assume that $n$ be even, if necessary by omitting one of the vertices.

We denote by \Ms{\all} the minimum stabbing number among all matchings
of $P$, by \Ts{\all} the minimum stabbing number of all spanning trees
of $P$, and by \TTs{\all} the minimum stabbing number of all
triangulations of $P$. 
We use  \Ms{2}, \Ts{2}, and \TTs{2} for the 
minimum axis-parallel stabbing numbers.

For a set $L=\{l_1,\ldots,l_n\}$ of line segments in the plane, the 
{\em crossing number} of a line $\ell$ is
the number of connected components of $\bigcup_{i=1}^n l_i\cap\ell$.
The {\em crossing number of} $L$ is the maximum crossing number over all
lines $\ell$; just like for the stabbing number,
the axis-parallel crossing number is defined for 
axis-parallel lines.
For matchings, trees, and triangulations, we use the analogous
abbreviations \CMs{\all}, \CTs{\all}, and \CTTs{\all}, and their
subscripted counterparts \CMs{2}, \CTs{2}, and \CTTs{2} for the
axis-parallel crossing numbers.  Note that stabbing and crossing
number coincide for planar matchings.

\section{Complexity}
\label{sec:complex}
In this section we prove \NP-hardness of computing the minimum stabbing number of
matchings and computing the minimum crossing number of triangulations; for spanning
trees the proofs are analogous, and we only give a sketch of the proof. Our
technique is rather general and should be applicable to other
structures and variants as well.

\subsection{Perfect Matchings}

\begin{theorem}
  \label{thm:MMSD2-5-NPC}
  Deciding whether $\Ms{2}\leq5$ is strongly \NP-complete.
\end{theorem}

\proof Clearly, the problem is in \NP. We show completeness using a
reduction from 3SAT~\cite{ml:GareyJohnson:79}.  Assume we have a
Boolean expression denoted by $B(x_{0},x_1,\ldots,x_{n-1})$ with $n$
variables and $k$ clauses of three literals each.  We construct a set
of vertices $P$ that has a perfect matching $M$ of stabbing number 5 if
and only if the Boolean expression can be satisfied; in case of an unsatisfiable 
expression, a stabbing number of at least 6 cannot be avoided.

\begin{figure}[htb]
   \begin{center}
   \input{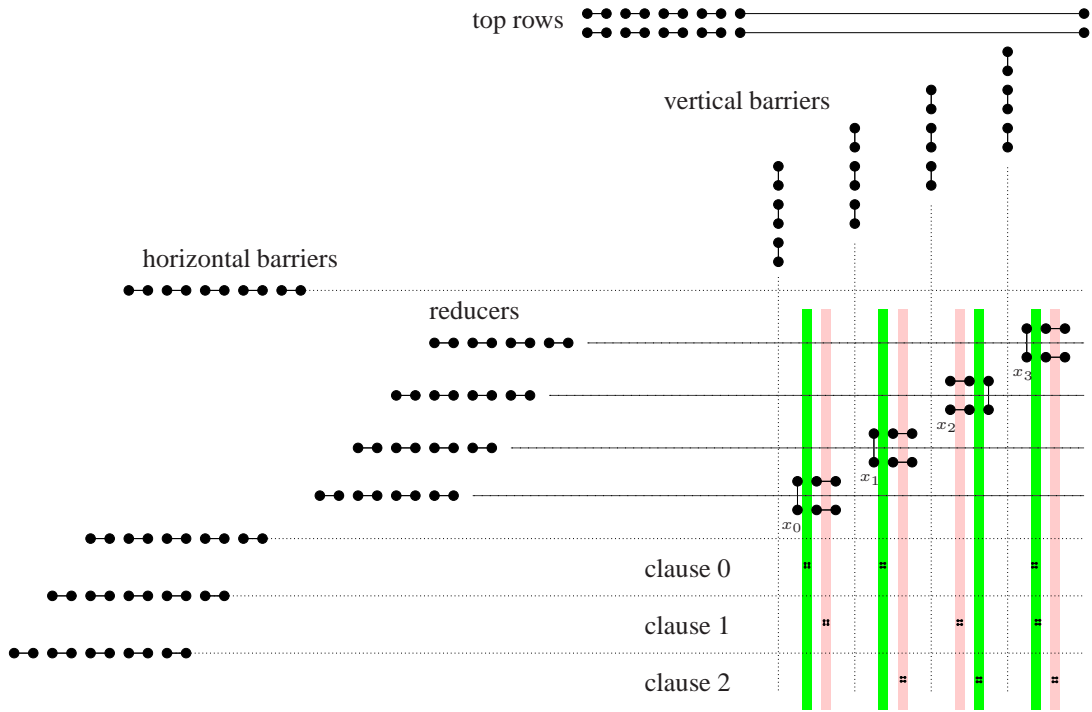}
   \caption{Overall layout of the construction for \Ms{2}. 
   Shown is the layout for the 3SAT instance 
   $(x_0\vee x_1\vee x_3)\wedge (\overline{x_0}\vee {x_2}\vee x_3)\wedge (\overline{x_1}\vee \overline{x_2}\vee \overline{x_3})$, with a truth setting of
$x_0=$ {\em true}, $x_1=$ {\em true}, $x_2=$ {\em false}, $x_3=$ {\em true}. Note that in full scale, the two gadgets representing the literal $x_3$ in clauses 
0 and 1 have disjoint $x$-coordinates.}
   \label{fi:mmsd5.0}
   \end{center}
\end{figure}

Consider the overall layout of $P$ as shown in Figure
\ref{fi:mmsd5.0}. We make critical use of the collinearity of vertices,
using up all of the available stabbing number of 5 in a particular
direction. Thus we are able to construct ``barriers'' which avoid any
interference between the different gadgets.

At the top of the layout are two groups of 10 vertices.  The vertices in a
group of 10 have the same $y$-coordinate.  We call these two groups
the \emph{top rows}.  The $i$-th vertex in the first top row has the
same $x$-coordinate as the $i$-th vertex in the second top row.  Below
the top rows are $n$ groups of 6 vertices. All vertices in a group of 6
have the same $x$-coordinate, as shown in the figure.  The vertical
lines through these groups of 6 vertices separate the variables from each other
and from other vertices left of the variables.  We call each
such group a \emph{vertical barrier} gadget.  The vertices in barrier $i$ are
sufficiently far below the vertices in barrier $i+5$ to ensure that horizontal
lines through the vertical barriers have stabbing number at most 5.
The barriers lie between vertical lines through the last and second last vertices of the top
rows.  To the left of the top rows and below the vertical barriers are
$k+1$ groups of 10 vertices.  Each vertex in a group of 10 has the same
$y$-coordinate.  
We call each such group a \emph{horizontal barrier} gadget.  
The vertices in barrier $i$ are sufficiently far to the right of
the vertices in barrier $i+5$ to ensure that vertical lines through the
horizontal barriers have stabbing number at most 5.  The horizontal
barriers are used to separate clauses from each other, to separate the
clauses from the variables, and to separate variables from other
vertices above the variables.  Between and to the right of the top two
horizontal barriers are groups of 8 vertices.  Each vertex in a group of
8 has the same $y$-coordinate.  We call each such group a
\emph{reducer} gadget of a variable.  The vertices in reducer $i$ are far
enough to the right of the vertices in reducer $i+1$ to ensure that
vertical lines through the reducers have stabbing number at most 5.

\begin{figure}[htb]
   \begin{center}
   \input{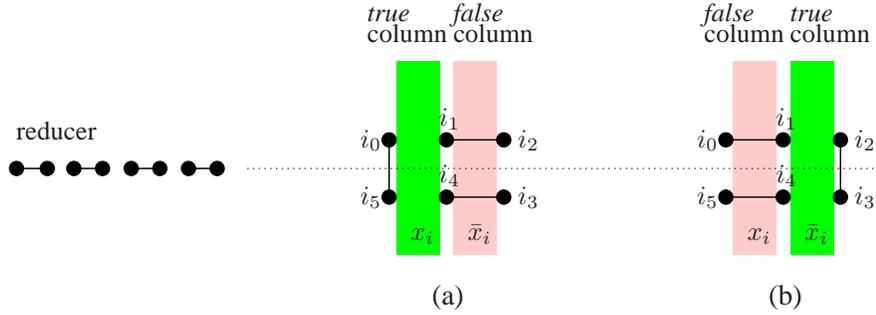}
   \caption{Variable $x_i$ with $x_i=$\;\emph{true} in (a) and
     $x_i=$\;\emph{false} in (b).} 
   \label{fi:mmsd5.1}
   \end{center}
\end{figure} 

Figure \ref{fi:mmsd5.1} shows a set of 6 vertices, numbered $i_{0}$ to
$i_5$ in clockwise order, that represents a variable.  These {\em variable}
gadgets are separated by vertical barriers, and are placed below the
vertical barriers, between the top two horizontal
barriers and to the right of the reducers.  
Variables are placed such that the $y$-coordinate of the
horizontal line through the reducer of a variable is in between the
$y$-coordinate of vertices $i_{0}$ and $i_5$ of that variable.
The strip formed by vertical lines between vertices $i_{0}$ and
$i_1$ of variable $x_i$ is called the \emph{$x_i$-column} of the
variable.  The strip formed by vertical lines between vertices $i_1$ and
$i_2$ of variable $x_i$ is called the \emph{$\bar {x}_i$-column} of
the variable. 

\begin{figure}[htb]
   \begin{center}
   \input{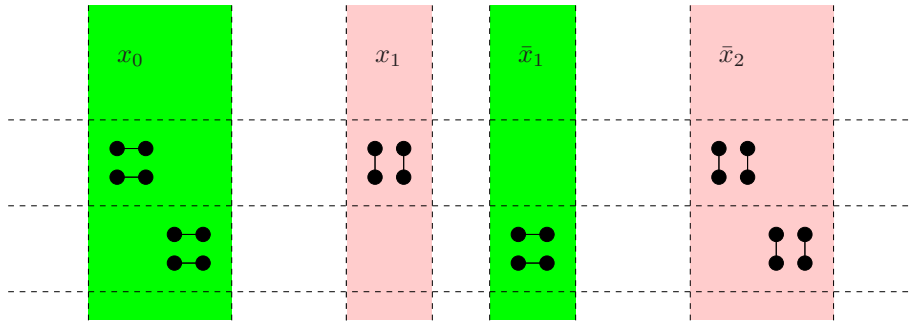}
   \caption{Clauses $(x_0 \vee x_1 \vee \bar {x}_2)$ and 
                    $(x_0 \vee  \bar {x}_1 \vee \bar {x}_2)$ with 
                    $x_0 = x_2 =$\;\emph{true} and $x_1 =$\;\emph{false}.}
   \label{fi:mmsd5.2}
   \end{center}
\end{figure}

The horizontal and vertical barriers create $kn$ locations for
{\em literal} gadgets. These are groups of 4 vertices representing the
occurrence of a variable in a clause.  Each group of 4 vertices forms an
axis-parallel square.  If a literal $x_i$ appears in the clause $c_j$,
we place a literal gadget in the $x_i$ column of clause $c_j$.  If a literal $\bar
{x}_i$ appears in the clause $c_j$, we place a literal gadget in the 
$\bar {x}_i$ column of clause $c_j$. The three literal gadgets for
a clause are put on the
same horizontal lines.  The literal in column $x_i$ of clause $c_j$ is
to the left of the literal in column $x_i$ of clause $c_h$ for $j <
h$.  Similarly, the literal in column $\bar {x}_i$ of clause $c_j$ is
to the left of the literal in column $\bar {x}_i$ of clause $c_h$ for
$j < h$.  Figure \ref{fi:mmsd5.2} shows the literals of two clauses.

We first assume that $B(x_{0},x_1,\ldots,x_{n-1})$ is satisfiable,
and show that $P$ has a matching $M$ of stabbing number 5.  We connect
vertex $i$ to vertex $i+1$ in each of the two top rows for
$i=0,2,4,6,8$.  We connect vertex $i$ to vertex $i+1$ in each vertical
barrier for $i=0,2,4$.  We connect vertex $i$ to vertex $i+1$ in each
horizontal barrier for $i=0,2,4,6,8$.  We connect vertex $i$ to vertex
$i+1$ in each reducer for $i=0,2,4,6$.

If the variable $x_i$ has the value \emph{true}, we connect the pairs
$(i_{0},i_5)$, $(i_1,i_2)$, and $(i_3,i_4)$ of the variable.  If the
variable $x_i$ has the value \emph{false}, we connect the pairs
$(i_{0},i_1)$, $(i_2,i_3)$, and $(i_4,i_5)$. Notice that if $x_i$ is
\emph{true}, then any vertical line in the $x_i$-column stabs 2 edges
in the top rows of $M$, and a vertical line in the $\bar {x}_i$-column
stabs 2 edges in the top row and 2 edges in the rectangle of the
variable.  If $x_{i}$ is \emph{false}, this situation is reversed.
The column with vertical stabbing number 2 is called the
\emph{true}-column 
of the variable, the column with vertical stabbing number 4 is
called the \emph{false}-column.  In each literal gadget representing the
value \emph{true} we connect the four vertices with two horizontal
edges.  In each literal gadget representing the value \emph{false} we connect
the four vertices with two vertical edges.

We can now verify that $M$ has stabbing number 5.  Any vertical line in
the \emph{true}-column of variable $x_i$ stabs two edges in the top rows
and at most two edges in a literal.  Any vertical line in the \emph{false}
column stabs two edges in the top rows, two edges of the variable and
at most one in a literal.  Any horizontal line in a clause stabs at
most three literal gadgets, one of which is set to \emph{true}. So these lines
stab at most 5 edges of $M$.  It can easily be verified that all other
horizontal and vertical lines stab at most 5 edges from $M$.

Conversely, we assume that $P$ has a matching of stabbing number 5.
We show that $B$ is satisfiable.  The matching used in this proof is
illustrated in Figures \ref{fi:mmsd5.0}, \ref{fi:mmsd5.1} and
\ref{fi:mmsd5.2}.  Because the top rows contain 10 vertices, these
vertices have to be connected to each other, otherwise the stabbing
number of $P$ exceeds $5$.  There are several ways to connect the sets
of 10 vertices. If we connect vertex $i$ to vertex $i+1$ for $i=0,2,4,6,8$
in each row, the number of edges stabbed by any horizontal or vertical
line is minimized. Therefore we may assume without loss of generality
that these edges are in the matching $M$.  Collinear vertices are
dealt with in a similar manner:
If there are several ways to connect a set of
collinear vertices, we will prefer connections
that have minimal stabbing number for all stabbing lines. 

Thus, we can connect vertex $i$ to vertex $i+1$ for $i=0,2,4,6,8$ in
each horizontal barrier gadget. For the same reason, we can connect vertex $i$
to vertex $i+1$ for $i=0,2,4,6$ in each reducer gadget.  Because vertical lines
through the vertical barrier gadgets stab two edges in the top rows, we can
connect vertex $i$ to vertex $i+1$ for $i=0,2,4$ in each vertical
barrier gadget. 

Now it is easy to see that no vertex in a variable 
or literal gadget for some variable $x_i$ can be matched with 
any vertex not involved with
representing the same variable $x_i$: otherwise, the edge
would cross a line through a vertical barrier, which already crosses
another five edges. Furthermore, such a vertex must be matched with
vertices from the same gadget (either the same variable gadget or the same 
literal gadget):
otherwise, we get a violation at a line through a horizontal barrier gadget.

Now each reducer gadget contributes four to a horizontal stabbing number. Thus, we
cannot connect the six vertices of a variable $x_i$ by three vertical
edges.  Figure \ref{fi:mmsd5.1} shows the two remaining, essentially
distinct matchings.

Of the three literals in each clause, one has to be set to true,
otherwise there will be a horizontal stabber intersecting six edges. The
true literal, say, $x$, must lie in a \emph{true}-column of a variable,
because vertical lines in this column have stabbing number two. Any other
literal in this column can also be set to \emph{true}. The literals
$\bar x$ lie in the \emph{false}-column of the same variable, and have
to be set to false. So if a matching of stabbing number five exists, there is a
truth assignment of the Boolean expression.  \qed

\begin{cor}
  There is no $\alpha$-approximation algorithm for $\Ms{2}$ with
  $\alpha<6/5$; in particular there is no polynomial time
  approximation scheme (PTAS), unless $\mathcal{P}=\NP$. 
\end{cor}

\begin{cor}
  \label{cor:MMSD_all-NPC}
  Computing $\Ms{\all}$ is a weakly \NP-hard problem.
\end{cor}

\begin{proof}
  We apply a perturbation technique, similar to the one in
  \cite{f-map-00}.  We start with the same basic
   construction as for the hardness proof
  for the axis-parallel case, and consider the grid formed by the
  coordinates of the resulting vertex set.  This grid is modified such
  that the interpoint distances between the vertices of the same gadget
  are 
  $\Theta(\eps^{n^2+2})$ for the literal gadgets and 
  $\Theta(\eps^{n^2})$ for all other gadgets. 
  Furthermore, the rest of the grid is
  perturbed by powers of $\eps$, such that only axis-parallel lines
  can stab more than two gadgets; in particular, we 
  increase the vertical distance
  between variable gadgets and the (narrower) literal gadgets 
  by a sufficient amount,
  in order to make sure that no line through two literal gadgets
  for the same variable can intersect the corresponding variable
  gadget. Now it is easy to see
  that lines that are not axis-parallel can stab at most four line segments,
  leaving only axis-parallel lines as critical.  \qed
\end{proof}

\subsection{Triangulations}
\label{subsec:triangles}

Our basic proof technique is the same as for matchings. We first describe
the construction of barrier gadgets, using the following terminology. A
\emph{horizontal line} is given by a set of vertices that are horizontally
collinear. A \emph{vertical line} is given by a set of vertically collinear
vertices.  A \emph{row} consists of two horizontal lines, and the (empty)
space between them. A \emph{column} consists of two vertical lines,
and the (empty) space between them.

\begin{lemma}
  \label{lem:tri_full_rows}
  Consider a row consisting of two horizontal lines $l_a$ and $l_b$ in $P$, having $a$
  and $b$ vertices, respectively.
If the combined number of edges on $l_a$ and $l_b$ is $a+b-i-2$,
then a horizontal stabber between $l_a$ and $l_b$
encounters at least $a+b+i-2$ triangles in any triangulation of $P$
and its crossing number is at least $a+b+i-1$.
\end{lemma}


\proof
Assume without loss of generality that $l_a$ lies above $l_b$.
Suppose there are $a-i_a-1$ edges on the line $l_a$
and $b-i_b-1$ edges on the line $l_b$ with $i_a + i_b = i$.
For each edge $(u,v)$ on $l_a$ there is a triangle $(u,v,w)$ where
$w$ lies either on or below $l_b$. Let $A$ denote this set of triangles.
Similarly, 
for each edge $(u,v)$ on $l_b$ there is a triangle $(u,v,w)$ where
$w$ lies either on or above $l_a$. Let $B$ denote this set of triangles.
For each two neighboring vertices $u$ and $v$ on $l_a$ for which there
is no edge $(u,v)$ there are triangles $(u,u_0,u_1)$ and $(v,v_0,v_1)$
such that $u_0$ and $v_0$ lie on or below $l_b$ and $u_1$ and $v_1$
lie above $l_a$. 
Let $I_a$ denote this set of triangles.
Also
for each two neighboring vertices $u$ and $v$ on $l_b$ for which there
is no edge $(u,v)$ there are triangles $(u,u_0,u_1)$ and $(v,v_0,v_1)$
such that $u_0$ and $v_0$ lie on or above $l_a$ and $u_1$ and $v_1$
lie below $l_b$.
Let $I_b$ denote this set of triangles. 
It is not hard to verify that any two of the the four sets 
of triangles $A$, $B$,
$I_a$ and $I_b$ have an empty intersection.
A horizontal line $l$ between
$l_a$ and $l_b$ stabs every triangle in $A$, $B$,
$I_a$ and $I_b$. So $l$ stabs at least $|A| + |B| + |I_a| + |I_b| $
$= (a-i_a-1 ) + (b-i_b-1) + 2i_a + 2i_b$ $= a + b + i - 2$ triangles 
and $ a + b + i - 1$ edges.
\qed

The lemma holds analogously for two vertical lines that form a column.  When a
row consists of two horizontal lines that have $\CTTs{2}+1$ vertices
altogether, we call it \emph{full} or fully triangulated. It follows from the lemma
that all $\CTTs{2}-1$ edges on the lines $l_a$
and $l_b$ have to be present.

\begin{theorem}
  \label{thm:Tri2-NPC}
  Finding \CTTs{2} is \NP-hard.
\end{theorem}

\begin{figure}[htb]
   \begin{center}
   \scalebox{.5}{\epsffile{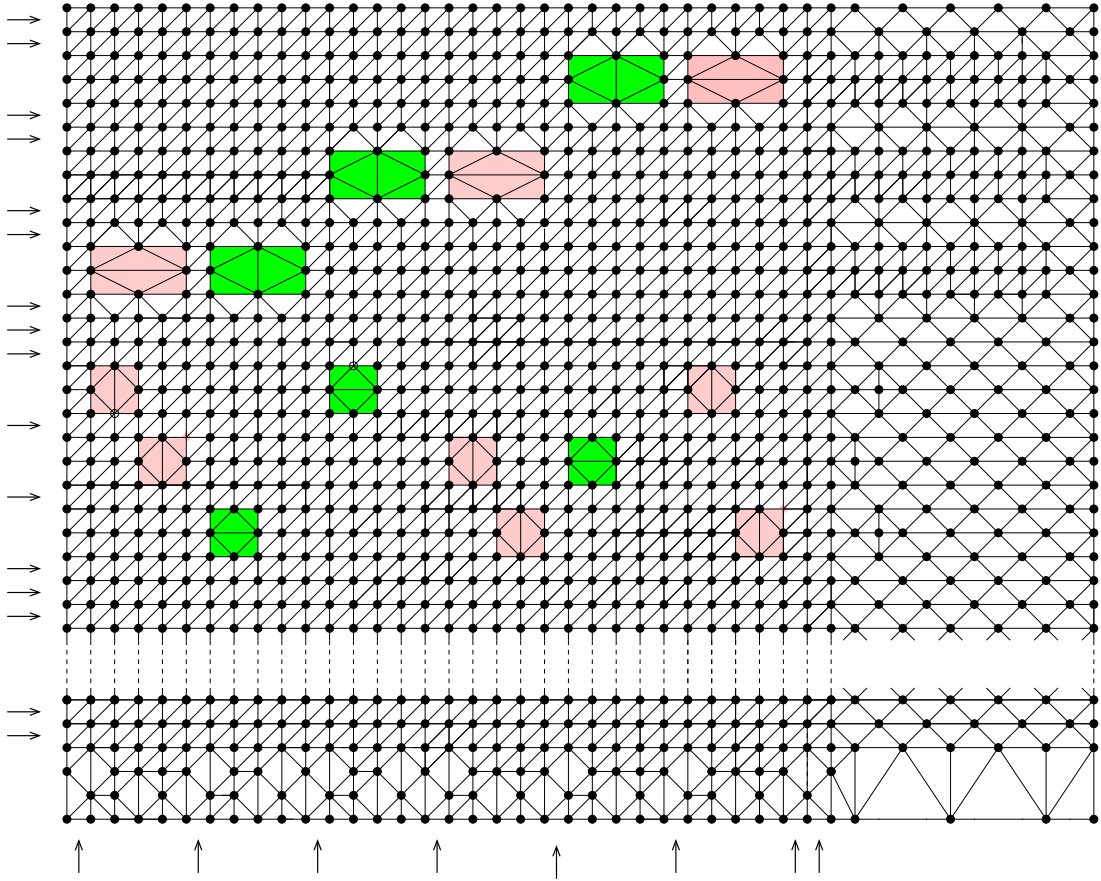}}
   \caption{Overall layout for \CTTs{2}. Clauses are $(x_{0} \vee x_1
     \vee \bar{x}_2)$, $(x_{0} \vee \bar{x}_1 \vee x_2)$, and
     $(\bar{x}_{0} \vee \bar{x}_1 \vee \bar{x}_2)$, with
     $x_{0}=$\;\emph{false} and $x_1=x_2=$\;\emph{true}.
     Arrows indicate full rows and columns, light or dark shading
     indicates true or false variables and literals as before.}
   \label{fig:triangles.nphard}
   \end{center}
\end{figure}

\begin{figure}[htb]
   \begin{center}
   \input{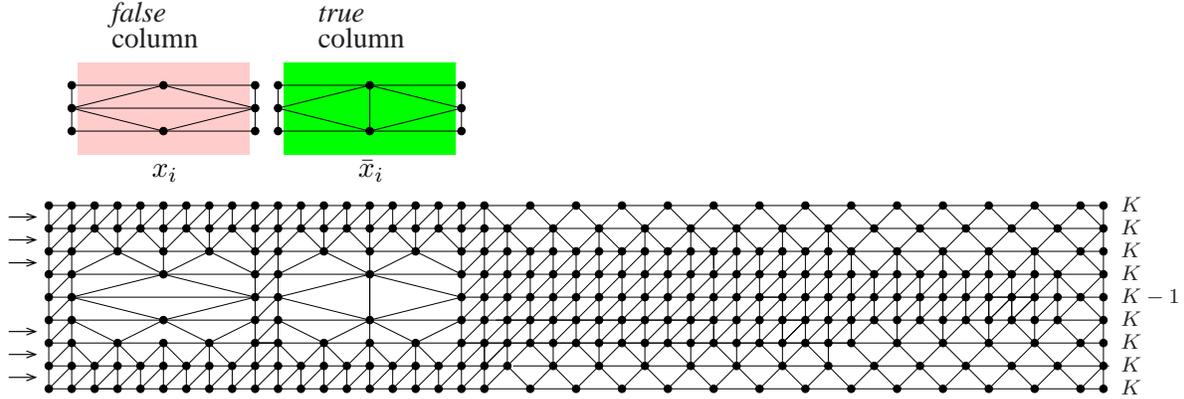}
   \caption{A variable gadget and how it is embedded in a grid of
     vertices.  Arrows indicate full rows.}
   \label{fig:variable.nphard}
   \end{center}
\end{figure}

\proof
Again we use a reduction from 3SAT, and the proof proceeds
along the lines of the proof of Theorem~\ref{thm:MMSD2-5-NPC}. 
See Figure~\ref{fig:triangles.nphard} for a
schematic layout of a representing vertex set $P$ for the 3SAT instance
$B(x_{~0},x_1,x_2) = (x_{0} \vee x_1 \vee \bar{x}_2) \wedge (x_{0}
\vee \bar{x}_1 \vee x_2) \wedge (\bar{x}_{0} \vee \bar{x}_1 \vee
\bar{x}_2)$.  Figure~\ref{fig:variable.nphard} shows the structure of
variable gadgets.

For a given Boolean expression $B(x_{0},x_1,\ldots,x_{n-1})$ with
$n$ variables and $k$ clauses of three literals each we construct a
set $P$ of vertices.  We show that there is a value $K$ such that $B$ is
satisfiable only if $\CTTs{2} = 2K-1$; if $B$ cannot be satisfied,
$\CTTs{2}$ is at least $2K$.

In Figure~\ref{fig:triangles.nphard} we have $K=39$ 
and a grid of vertices with some well-defined 
holes.  The maximum number of vertices in a horizontal or
vertical line is $K$, and many lines have exactly
$K$ vertices.  By Lemma~\ref{lem:tri_full_rows} a full row or column in
this setting has exactly $2K-2$ triangles.  
Gadgets are separated by full rows and columns.

Figure~\ref{fig:variable.nphard} shows two horizontally aligned
rectangles of eight vertices each that together represent a variable
$x_i$. We call the strip formed by vertical lines that stab the left
rectangle the $x_{i}$-{\em column}, and the strip formed by vertical lines that
stab the right rectangle the $\bar{x}_{i}$-{\em column} of the variable.
The gadget works essentially the same way as that in
Figure~\ref{fi:mmsd5.1}. Each rectangle has
full rows and columns as neighbors. We indicate how this can be
achieved in horizontal direction in Figure~\ref{fig:variable.nphard}.
By Lemma~\ref{lem:tri_full_rows} we conclude that all edges on the boundary 
of the convex hull of each rectangle are present in any triangulation of
minimal crossing number. The horizontal lines that contain the top and
bottom lines of the two rectangles, respectively, contain $K$ vertices
each; the horizontal line that passes through the middle 
of the rectangles contains $K-1$ vertices. 
Therefore Lemma~\ref{lem:tri_full_rows}
shows that exactly one
edge along this horizontal line in the middle of the rectangle may be missing.
We call the strip that is spanned by all
vertical lines that stab the rectangle with the missing horizontal edge the
\emph{true}-column of the variable $x_i$.  The strip that is spanned
by all vertical lines that stab the rectangle for which the 
middle horizontal edge
is present is the \emph{false}-column of the variable $x_i$.  As shown in 
Figure~\ref{fig:triangles.nphard}, any satisfying literal
adds exactly one less to the vertical crossing 
number than an unsatisfied one. In the overall layout, variable
$x_{i}$ is placed below and to the left of variable $x_{j}$ for $i<j$ in
such a way that variables are vertically separated from one another by
a full row, and horizontally separated from one another by a full
column.

Each literal is represented by a square with eight vertices on its boundary. 
We make the width
of rectangles of the variables equal to a power of two,
and wide enough to accommodate the necessary
number of literals. That is, each rectangle of a variable is of a
width at most four times the number of 
occurrences of the most frequent literal in $B$. 
Figure~\ref{fig:variable.nphard} gives a hint at how this widening of
a rectangle is done. Notice that if the most frequent literal
in $B$ occurs $t$ times, there are $\Theta(\log t)$ rows
above and below the variable so that in the top and bottom lines 
of these groups of rows
the vertices above and below the variables are a distance of one apart.
The three literals of a clause $c_j$
are horizontally aligned, and the two rows that are spanned by them
are called the clause $c_j$. 
Clauses are separated from each other by full rows.
If a literal $x_i$ appears in the clause
$c_j$, we place a literal gadget in the $x_{\,i}$-column of the clause
gadget $c_j$.  If a literal $\bar {x}_i$ appears in the clause $c_j$,
we place a literal gadget in the $\bar {x}_{\,i}$-column of the clause
gadget $c_j$. The literal in column $x_{\,i}$ of clause $c_j$ is to the
left of the literal in column $x_{\,i}$ of clause $c_h$ for $j < h$.
Similarly, the literal in column $\bar {x}_{\,i}$ of clause $c_j$ is to
the left of the literal in column $\bar {x}_i$ of clause $c_h$ for $j
< h$. So 
no vertical line stabs the interior of more
than one literal.

By adding vertices to the right of the literals we ensure that a
horizontal line through the top or bottom row of the three literals of
a clause has exactly $K$ vertices, and a horizontal line through the
middle horizontal line of the three literals of a clause has exactly $K-2$
vertices. So two edges along these middle lines may be missing in
a triangulation of minimum crossing number, but no more than two. As we
will argue later, these edges will be missing in the interior of at most two
of the literals. We call the missing of the horizontal middle edge in
a literal the \emph{false} setting of the literal, and the presence of
this edge within a literal the \emph{true} setting of the literal.

Because the rows above and below a variable are full, we can assume 
that they are triangulated as shown in 
Figure~\ref{fig:variable.nphard}, because any other triangulation would
result in strictly higher vertical crossing numbers.
So the  \emph{true} column of a variable has a vertical crossing number that 
is one less than the crossing number of a \emph{false} column.
In adding vertices at the bottom of the clauses we ensure the
following vertical vertex counts. First of all, the columns neighboring
the variables have to be full; in particular, vertical lines that
stab the left or the right vertices of a variable rectangle have $K$
vertices each.  
All other vertical lines through a variable 
should get a number of vertices so that 
if the corresponding column of the variable is set to \emph{false}, 
all remaining edges on these vertical lines have to be present. 
This implies that if the column is set to \emph{true}, 
we can have one missing edge
in the vertical line to the right of the left boundary
of the rectangle, and if this edge is missing, there cannot be
a missing edge in the next vertical line, 
one edge missing in the next line, etc. 

Let $B(x_{0},x_1,\ldots,x_{n-1})$ be satisfiable. We show that $P$ has a
triangulation of crossing number $2K-1$ that is minimum by
Lemma~\ref{lem:tri_full_rows}. All full rows and full columns are
fully triangulated. 
If variable $x_{i}$ has the value \emph{true}, we triangulate the
interior of the two rectangles of variable $x_{i}$ in such a way that
the $x_{i}$-column becomes this variable's \emph{true}-column, and the
$\bar{x}_{i}$-column becomes this variable's \emph{false}-column. The
triangulation of the interior of the rectangles is reversed when
variable $x_{i}$ has the value \emph{false}. We set each literal that
represents the value \emph{true} to its \emph{true} setting, and set
each literal that represents the value \emph{false} to its
\emph{false} setting. The triangulation can be completed arbitrarily. 

We can now convince ourselves that such a triangulation of $P$ indeed
attains $\CTTs{2}=2K-1$. No fully triangulated row or column has a
crossing number larger than $2K-1$. 
%
%
Because exactly one edge is missing
in the horizontal middle line of each variable,
Lemma \ref{lem:tri_full_rows} implies that
the crossing numbers of the two rows of a variable are both equal to
$(2K-1) + 1 - 1 = 2K-1$.
In each clause there is at least one literal
in its \emph{true} setting. Therefore, we can afford two 
extra triangles caused by the \emph{false} setting of the other two literals
in the clause, and no row of a clause intersects more than $2K-2$
triangles. Finally, we may have one edge missing on the vertical lines
passing trough the middle of a literal only in a \emph{true}-column.
This condition holds by definition of the setting of the literals
according to the truth value of the variables. The ``arbitrary
completion'' of the triangulation only happens in the lower right of
the construction. In this corner, vertical and horizontal lines have a 
low vertex count (except for the boundary), and the allowed crossing 
number is not exceeded.

For seeing the converse, assume that there is a triangulation of $P$ that
has crossing number $2K-1$. We show that $B$ is satisfiable.  Because a
full row or column can be triangulated in such a way that the crossing
number of $2K-1$ is not exceeded, we only have to take care of the rows
and columns in which we have a degree of freedom, and where the vertex
count is critical by Lemma~\ref{lem:tri_full_rows}.  
%
%
Because 
both the top row and the bottom row of a clause have $2K-2$ vertices, 
we can
afford at most two literals of each clause set to \emph{false}.  One
literal in each clause has to be set to \emph{true}. The \emph{true}
literal has to be in a \emph{true}-column of a variable, for otherwise
the vertical crossing number would exceed $2K-1$. Any other literal in
this column can also be set to \emph{true}. The horizontal vertex count
of the horizontal lines of the variable forces the second column of
this variable to be a \emph{false}-column. In order not to exceed the
allowed crossing number in vertical direction, all literals in this
column have to be set to \emph{false}. As one easily checks, this
yields a consistent setting of the variables, and $B$ is satisfiable.

Finally, the size of our construction is indeed polynomial:
let $n$ and $c$ be the number of variables and clauses
of $B(x_{0},x_1,\ldots,x_{n-1})$. Let $t$ be the number of times the most
frequent literal in used in $B$.
A rectangle
that represents  a variable has width at most $4t$
and requires $\Theta(\log t)$ rows. 
%
%
The number of rows used by 
the clauses is $\Theta(c)$.
In order to achieve the correct vertex count 
in each line 
we may have to add vertices to the right and below,
which requires at most $\Theta(t)$ additional rows and columns.
Therefore $K$ is 
polynomial in $n$, $c$ and $t$.
\qed

\subsection{Spanning Trees}
\label{sec:trees}

\begin{figure}[hbtp]
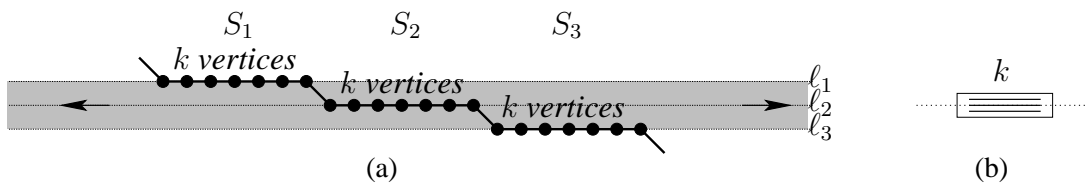

  \centering
  \input stair.pstex_t
  \caption{A horizontal barrier gadget, consisting of $3k$ vertices: 
    (a) In a spanning tree of stabbing number $k+1$,  
    no edge may cross the shaded region. (b) Symbol for the
    barrier gadget; the dotted line indicates the blocked strip.}
  \label{fig:stair}
\end{figure}

The basic construction for showing hardness of finding a spanning tree
of minimum stabbing number is similar to the one for matchings.
As before, we use barriers to restrict possible connections: we make
use of the arrangement shown in Figure~\ref{fig:stair}, which works as
a barrier gadget because of the following lemma.

\begin{lemma}
  \label{lem:stair}
  Consider three parallel lines, $\ell_1$, $\ell_2$, $\ell_3$, with
  a set $S_i$ of $k$ vertices on line $\ell_i$, $i=1,2,3$; 
  let $S=S_1\cup S_2\cup S_3$. Consider $P\supset S$ and a
  spanning tree $T$ of $P$ with stabbing number $k+1$. Then
  no edge of $T$ crosses the strip spanned by the three
  lines.
\end{lemma}

\begin{proof}
See Figure~\ref{fig:stair}.
Consider a spanning tree $T$ of $P$, with $v\in P\setminus S$ lying outside
of the strip. Orient all edges of $T$ towards $v$. Each vertex in $S$
must have outdegree 1, meaning that there are $k$ outgoing edges
for each of $S_1$, $S_2$, $S_3$, contributing $k$ to the stabbing
numbers along $\ell_1$, $\ell_2$, $\ell_3$. 
One of the outgoing edges of $S_2$ must intersect
$\ell_1$ or $\ell_3$ in order to connect $S_2$ to the 
rest of the graph; thus, one of those two lines stabs
$k+1$ edges, implying the claim.
\qed
\end{proof}

\begin{figure}[hbtp]
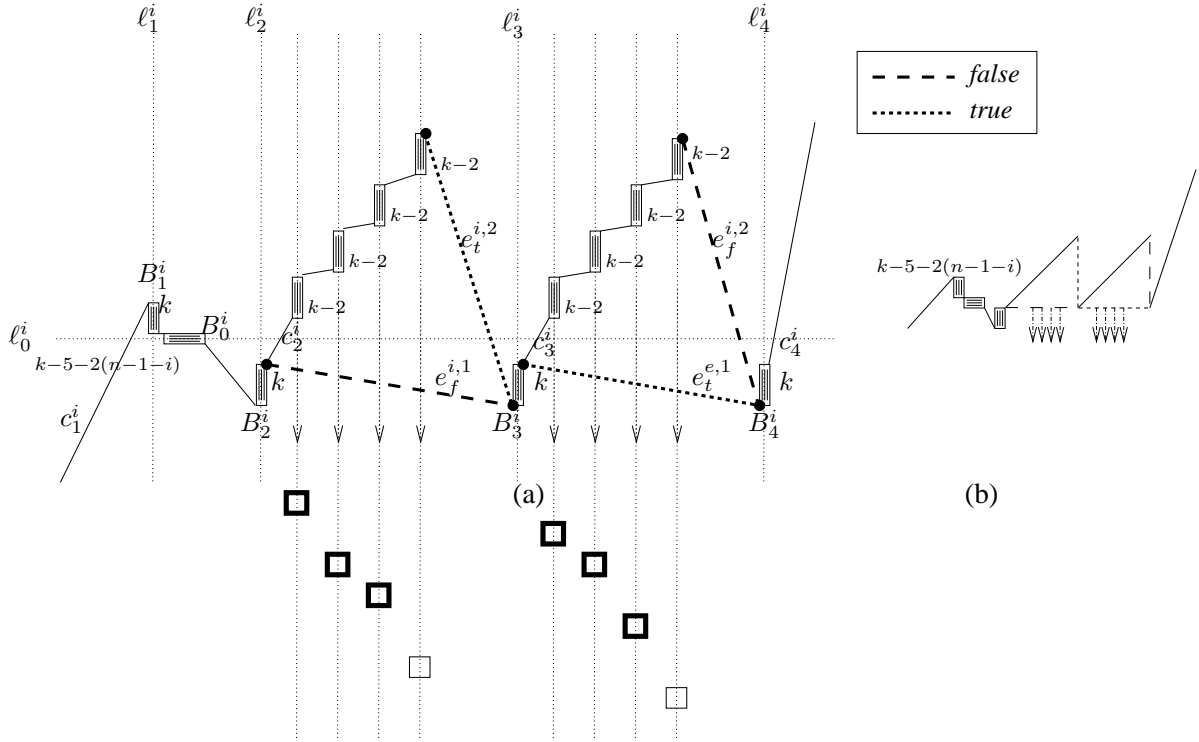

  \centering
  \input variable2.pstex_t
  \caption{A variable gadget for variable $x_i$. (a) In a spanning tree of stabbing number $k+1$, 
  the \emph{true} or the \emph{false} setting is chosen.
   (b) Symbol for the variable gadget.}
  \label{fig:variable}
\end{figure}

Our variable gadgets look as in Figure~\ref{fig:variable}; shown is
the gadget for variable $x_i$; note that the gadgets for $x_0,\ldots,x_{i-1}$
are left and below the box spanned by the gadget, while
the gadgets  
for $x_{i+1},\ldots,x_{n-1}$ are above and to the right of the spanning box.
The bold squares below the gadget indicate the position of literal
gadgets, which wil be discussed further down.

Also note the use of vertical barrier gadgets: 
a number of $i$ next to it indicates
a gadget consisting of $3i$ vertices, which alread requires a crossing
number of $i+1$; thus, only $k-i$ additional edges in a spanning tree
may cross the dotted line induced by such a gadget. 
The arrows pointing down from the 
bottom indicate a number of literal gadgets, consisting of 
$2\times 2$ arrangements of vertices. See Figure~\ref{fig:trees.npc}
for the resulting overall arrangement.

\begin{lemma}
  \label{lem:variable}
  Let $S$ be the arrangement of vertices shown in
  Figure~\ref{fig:variable}, with barrier gadgets placed and sized as
  indicated, and let $P\supseteq S$.  Let $P$ be constructed
  as shown in Figure~\ref{fig:trees.npc}. Then any spanning tree of $P$
  that has stabbing number at most $k+1$ must use at least one of the two
  edges at the bottom of the arrangement, labeled 
  $e_t^{i,1}$  (for \emph{true}) or $e_f^{i,1}$ (for \emph{false}.)
\end{lemma}

\begin{proof}
Assume there is a spanning tree of stabbing number at most $k+1$.
Consider the barrier gadgets labeled $B^i_1$, $B^i_2$, $B^i_3$, $B^i_4$,
and the corresponding lines, $\ell^i_1$, $\ell^i_2$, $\ell^i_3$, $\ell^i_4$.
By the previous lemma, no edge can cross one of those lines.
Therefore, the literal boxes below each clause must be connected within
the vertical strips bounded by $\ell^i_2$ and $\ell^i_3$ on one hand, or
$\ell^i_3$ and $\ell^i_4$, on the other hand. This requires at least one
edge within each of the two strips to cross the line $\ell^i_0$. Moreover,
the lines $\ell^i_1$ and $\ell^i_4$ must not be crossed,
implying that the variable gadgets are connected to their neighbors
at barriers $B^i_1$ and $B^i_4$. These connections between neighboring variable
gadgets form a stair-like chain of variable gadgets,
as shown in Figure~\ref{fig:trees.npc}: connections are in 
increasing $x$- and $y$-order, which correspond to
increasing variable indices; we call this the {\em exterior stair}.
Similarly, the barrier gadgets associated with each of the two truth 
settings of a variable form one stair-like chain each, 
as shown in Figure~\ref{fig:variable}; we call these the 
two {\em interior stairs}.

Now consider the horizontal
barrier $B^i_0$, consisting of three groups of $k-5-2(n-1-i)$ vertices each. 
By the previous arguments, the line $\ell_0$ has to cross all of the edges
connecting the true and false literal boxes of the $n-1-i$ 
variables with higher indices,
i.e., cross $2(n-1-i)$ edges. 
Furthermore in variable $x_i$ there are four other edges that connect vertices
above line $l^i_0$ to vertices below $l^i_0$ and so are crossed by $l^i_0$.
For example these edges could be the ones
labeled $c^i_1,c^i_2,c^i_3,c^i_4$ in the figure.
This allows only one of the edges $e_t^{i,2}$ and $e_f^{i,2}$ to be
used for connecting the two interior stairs
with each other and with the exterior stair; thus, at least one of the edges 
$e_f^{i,1}$ and $e_t^{i,1}$
must be used, proving the claim.
\qed
\end{proof}

Making use of the above gadgets, we get

\begin{theorem}
  \label{th:treehard}
  It is \NP-hard to determine \Ts{2}. 
\end{theorem}

\begin{proof}
The basic idea for the construction is similar to the one used in the previous 
sections, making use of Lemmas~\ref{lem:stair} and \ref{lem:variable}.  
The use of gadgets and the overall layout of the construction are shown
in Figure~\ref{fig:trees.npc}. 

\begin{figure}[hbtp]
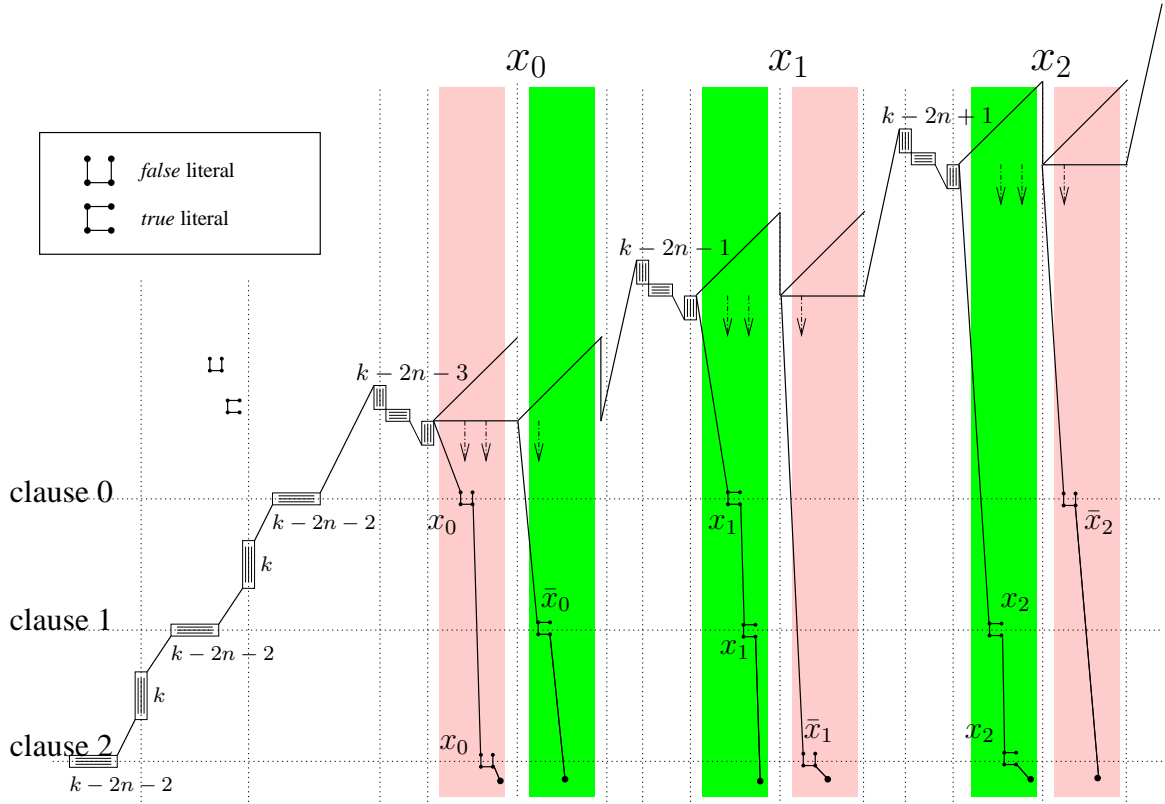

  \centering
  \input npc.tree.pstex_t
  \caption{The overall layout for the hardness proof for spanning trees.
    There is a total of $n$ variables; $k$ is a sufficiently large
    number. In question is the existence of a spanning tree with
    stabbing number $k+1$.  Shown is the representation of the 3SAT
    instance $(x_0\vee x_1\vee \bar{x}_2)\wedge (\bar{x}_0\vee x_1\vee
    {x}_2)\wedge (x_0\vee \bar{x}_1\vee {x}_2)$, for $n=3$, with
    $x_0=$ \emph{false} and $x_1=x_2=$ \emph{true}. }
  \label{fig:trees.npc}
\end{figure}

Given a Boolean expression
$B(x_{0},x_1,\ldots,x_{n-1})$, we can find a vertex set that has a
spanning tree of a stabbing number $k+1$, if and only if $B$ is
satisfiable.  Consider the vertex
set as given in Figure~\ref{fig:trees.npc} with $k=3n$. If
$B(x_{0},x_1,\ldots,x_{n-1})$ has a satisfying truth assignment, we first connect
the vertex on the left side of the drawing into one long path. This
path contains all horizontal and vertical barrier gadgets as shown in
Figure \ref{fig:trees.npc}.  We connect the variable and literal
gadgets according to their values in $B$.  In each \emph{true}- and
\emph{false}-column, there is a vertex lower than all horizontal
barriers, and to the right of all literals in that column. We connect
this vertex to the right lowest vertex of the lowest literal in the same
column.  Because each clause has at least one true literal, the
horizontal stabbing number of a stabber through a clause is at most
$k$, because the horizontal barrier on the left contributes $k-2n-1$,
the variable(s) set to true and the $n-3$ variables that do not appear
in the clause contribute two each, and the variable(s) set to false
contribute three each, which is at most $k-2n-1+2(n-2)+6 = k+1$.  A
horizontal stabber through the horizontal barrier in the $i$-th
variable gadget stabs $k-5-2(n-1-i)+1$ edges in the barrier, one more
edge to the left of the barrier, four edges of the variable to the
right of the gadget and two more for each of the $n-i$ variables to
its right, for a total of $k-5-2(n-1-i)+1+5+2(n-1-i) = k+1$. A vertical
stabber through literals stabs at most $k-1$ edges in the vertical
barriers, plus two more, either one from a false literal and one in
the variable gadget, or two edges in the true literals.  So we have a
spanning tree of stabbing number $k+1$.

Conversely, assume that the vertex set has a spanning tree of stabbing 
number $k+1$. As we showed in the proof of Lemma~\ref{lem:variable}, 
the literal gadgets can only be connected within their respective
strips, forcing at least $2n$ stabbed edges. Furthermore, each
\emph{false} literal (being connected in a ``u''-like fashion)
causes an additional stabbed edge, while a \emph{true} literal
(connected in a ``c''-like fashion) does not cause any additional
stabbings. Thus, each clause must contain at least one \emph{true}
literal. Because of Lemma~\ref{lem:variable}, at least one of the
edges $e_f^1$ and $e_t^1$ must be present, guaranteeing that
only the negated or only the unnegated literals for each variable
can be connected in a ``c''-like fashion, i.e., forcing a feasible
setting of the variables. Thus, we get a truth setting of the variables
that satisfies $B$.
\qed
\end{proof}

This immediately implies

\begin{cor}
  \label{cor:Cr-treehard}
  It is \NP-hard to determine \Ts{\all}. 
\end{cor}

\begin{proof}
The argument is similar to the one in
Corollary \ref{cor:MMSD_all-NPC} for matchings:
use the construction of Theorem~\ref{th:treehard}, for which
the criticality of certain axis-parallel lines requires
satisfying a 3SAT instance in order to achieve low stabbing number.
Scale down the bounding boxes for all gadgets, with
literal gadgets ending up in appropriately smaller bounding boxes.
Then perturb the position of the gadgets, shifting all
vertices in the same gadget by the same amount, such that no line can intersect
the bounding boxes of any three gadgets, again inserting a sufficient
vertical distance for excluding a line that stabs a variable gadget and 
two of its literal gadgets.  This leaves only
the axis-parallel lines to be critical, implying the same combinatorial
behavior as in the axis-parallel case.
\qed
\end{proof}

\begin{figure}[hbtp]
  \centering
  \scalebox{.5}{\epsffile{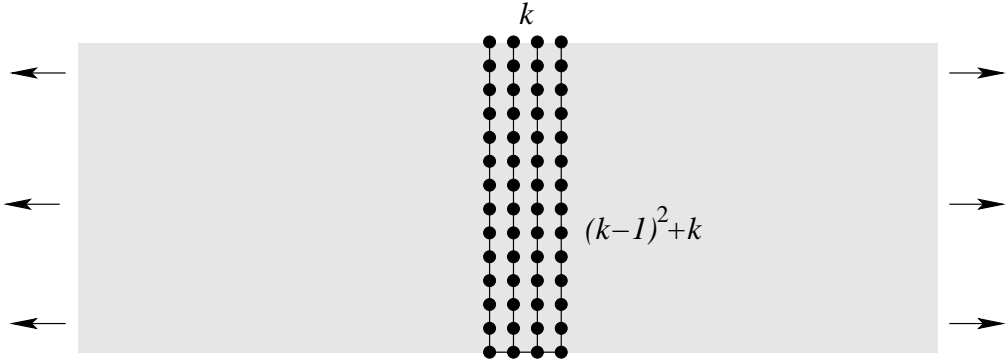}}
  \caption{A barrier gadget for showing hardness of minimizing the crossing number of a spanning tree.}
  \label{fig:stair.cross}
\end{figure}

The hardness proof for minimizing the crossing number has the same
structure as the one for stabbing number. Instead of the barrier gadget
implied by Lemma~\ref{lem:stair}, we use a slightly different one,
as shown in Figure~\ref{fig:stair.cross}.

\begin{lemma}
  \label{lem:stair.cross}
  Let $S$ be the $k\times ((k-1)^2+k)$ arrangement of vertices shown in 
  Figure~\ref{fig:stair.cross}, and let $P\supseteq S$.
  If $P$ has a spanning tree $T$ with crossing number $k$, then
  no edge of $T$ connecting two vertices outside of the arrangement
  crosses the horizontal strip spanned by the arrangement.
\end{lemma}

\begin{proof}
Suppose there is a crossing edge. Consider the $(k-1)^2+k$
horizontal lines that pass through $k$ vertices of the vertex arrangement.
Because of the crossing edge, the intersection of each line
with the edges and vertices within each corresponding horizontal set 
cannot consist of more than $k-1$ connected components, so
this intersection must contain at least one horizontal edge within the set; 
thus, a total of at least $(k-1)^2+k$ horizontal edges within
the arrangement cannot be avoided.
Now consider the $k$ vertical lines that pass through $(k-1)^2+k$
vertices of the arrangement; these lines subdivide the plane into 
$k+1$ vertical strips, $k-1$ of which have width 1. 
Any of the at least $(k-1)^2+k$ horizontal edges within
the arrangement must cross at least one of the intermediate strips; thus,
the average number of horizontal edges per bounded strip
is at least $\frac{(k-1)^2+k}{k-1}=(k-1)+\frac{k}{k-1}>k$. 
By the pigeonhole principle, this implies that there must be a bounded vertical
strip that is crossed by more than $k$ edges, a contradiction to
our assumption that the crossing number is at most $k$.
\qed
\end{proof}

As the figure shows, there is a feasible subtree of the gadget, as long
as no edge crosses the indicated strip. This allows us to use the 
arrangement as a barrier gadget. As for the rest of the construction
for the proof of Theorem~\ref{th:treehard}, stabbing and crossing numbers
coincide, this immediately implies the following.

\begin{theorem}
  \label{thm:Cr-Tree-NPC}
  It is \NP-hard to determine \CTs{2}. 
\end{theorem}

 From this it is easy to derive the following, again using a perturbation argument.

\begin{cor}
  \label{cor:treehard}
  It is \NP-hard to determine \CTs{\all}. 
\end{cor}

\begin{proof}
The argument is similar to the one in Corollary \ref{cor:Cr-treehard}:
use perturbation to ensure that only axis-parallel lines can stab more than
two gadgets.
The only additional difficulty that has to be overcome is the fact
that a diagonal line through one of the barrier
gadgets in Figure~\ref{fig:stair.cross}
may have crossing number $k$: in principle, crossing number $2k$ could
arise from stabbing two such gadgets in a non-axis-parallel manner.
However, scaling the bounding boxes of the barrier gadgets in a way that
the vertical dimension is much smaller than the horizontal dimension
(say, by a factor of $O(\eps^{n^3})$) 
makes sure that only lines of slope within range $[-\eps^{n^3},\eps^{n^3}]$
achieve crossing number $k$ for one gadget. Thus, only 
almost horizonal lines are of concern; it is easy to see
that the above perturbation guarantees
that no such line intersects more than one barrier gadget.
Therefore, only axis-parallel lines can be critical, and the hardness
proof remains valid.
\end{proof}


\section{Integer Linear Programs for Minimum Stabbing Number}
\label{sec:LP}

In view of the negative complexity results for our problems there are
two major directions to proceed: providing (good) lower bounds on the
minimum stabbing number in order to obtain approximation algorithms;
and insisting on optimality despite \NP-hardness. Our (integer) linear
programming approach is an elegant way to combine both issues. We deal
with them in the next two sections.

\subsection{Perfect Matchings}

In combinatorial optimization, $P$ corresponds to the vertex set
of a straight-line embedded complete graph $G=(P,E)$; then a
matching $M$ can be represented by its edge incidence vector
$x\in\{0,1\}^E$, where $x_{ij}=1$ if $ij\in M$, and $x_{ij}=0$
otherwise. Using these variables we are able to state an integer linear 
program for finding a perfect matching of minimum stabbing number. For
$S\subseteq P$, denote by $\delta(S)=\{ij\in E\mid i\in S, j\notin
S\}$ the \emph{cut} induced by $S$.
\begin{alignat}{3}
  &\text{minimize} & \enspace k\label{eq:objk}\\
  &\text{s.t.}&\sum_{ij\in\delta(\{i\})}x_{ij}  &=  1 &&\qquad
  \forall\, i\in P 
  \label{eq:degree}\\ 
  &&\sum_{ij\in\delta(S)}x_{ij} & \geq 1 &&\qquad \forall\, 
  \emptyset\neq S\subset P,\; |S| \text{
    odd}\label{eq:blossom}\\
  &&\sum_{ij : ij \cap \ell(d) \not= \emptyset} x_{ij} &\leq k
  &&\qquad \forall \text{ stabbing line $\ell(d)$ in
    direction } d  \label{eq:stab_constraints}\\
  &&x_{ij} & \in \{0,1\}&&\qquad
  \forall\, ij\in E \label{eq:integ}\\
  \intertext{We obtain the associated linear programming (LP)
    relaxation by replacing \eqref{eq:integ} by} &&x_{ij} & \geq
  0\tag{\ref{eq:integ}$'$}\label{eq:nonneg}
\end{alignat}
The inequalities~\eqref{eq:degree} and~\eqref{eq:blossom} are
necessarily satisfied for any perfect matching given by $x$,  
where the {\em blossom constraints}~\eqref{eq:blossom} ensure that any 
subset of vertices of odd cardinality has at least one edge to the outside.
In his seminal paper Edmonds~\cite{ml:Edmonds:65} showed that---together
with~\eqref{eq:nonneg}---these inequalities already constitute the
complete description of the perfect matching polytope, that is, its
extreme vertices exactly correspond to the incidence vectors of perfect
matchings in $G$, i.e., the LP relaxation is integral.

\begin{figure}[hbtp]
  \centering
  \scalebox{.31}{\epsffile{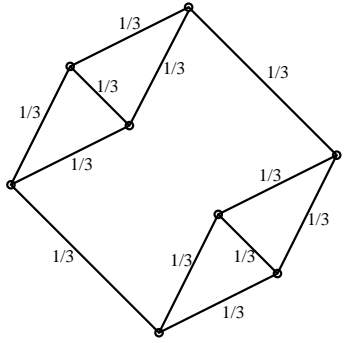}}
  \caption{An optimal fractional solution of value 4/3 with maximum edge 
weight $1/3$.}
  \label{fi:third}
\end{figure}

To this description we add the stabbing
constraints~\eqref{eq:stab_constraints}, for which we count the number of
intersections of matching edges with any given line; 
this number is
bounded by the variable $k$, and $k$ is minimized. We have to choose
this way of modeling because of our min-max objective. An optimal
solution $x$ to the {integer} program
\eqref{eq:objk}--\eqref{eq:integ} represents a matching with stabbing
number exactly \Ms{}. In a pure integer programming description,
the blossom constraints are implied; however, 
Figure~\ref{fi:third} shows that the introduction
of stabbing constraints yields a polytope that is no longer integral,
which is to be expected for an \NP-hard problem. That is why we
consider the LP relaxation, for which the 
use of blossom constraints yields a considerably tighter set of solutions.
We will make use of this fact in the following section; here we comment
on the complexity of the given LP and its solution.

We can limit the number of stabbing 
constraints~\eqref{eq:stab_constraints}: for the 
axis-parallel stabbing number, we can assume without loss
of generality that a critical stabbing line
runs through a vertex, so we get at most $O(n)$
restrictions for the axis-parallel stabbing number; for the 
general stabbing number, we may assume that a critical stabbing
line runs through two vertices, so we can limit ourselves to
$O(n^2)$ restrictions.
On the other hand, we
have exponentially many blossom inequalities;  
however, it is well 
known that blossom inequalities can be separated in polynomial 
time~\cite{PadbergRao}, i.e.,
one can check in polynomial time whether a given $x$ violates 
some blossom inequality,
and if so,  identify such an inequality.
This polynomial-time \emph{separation} allows us to solve the linear
programming relaxation \eqref{eq:objk}--\eqref{eq:nonneg} in strongly 
polynomial time~\cite[Thm.\ 5.11]{ml:Schrijver:03} by
means of the ellipsoid method \cite[Thm.\ 66.5]{ellipsoid}.  An optimal solution
$x$ will in general be fractional, and we speak of \emph{fractional
  stabbing number} in this context.  It is a lower bound on \Ms{}.

\subsection{Spanning Trees}

There are several polynomial-size LP formulations for spanning trees,
see \eg \cite{ml:MagnantiWolsey:95}.  However, similar to matchings,
we choose an exponential-size integer program that is again based on
cut constraints. 
We directly state the LP relaxation.
\begin{alignat}{3}
  &\text{minimize} & \enspace k\label{eq:tree_objk}\\
  &\text{s.t.}&\sum_{ij\in E}x_{ij} & =  n-1 \label{eq:total}\\
  &&\sum_{ij\in\delta(S)}x_{ij} & \geq  1 &&\qquad \forall \emptyset \neq S\subset P\label{eq:cut}\\
  &&\sum_{ij : ij \cap \ell(d) \not= \emptyset} x_{ij} &\leq k
  &&\qquad \forall \text{ stabbing line $\ell(d)$ in
    direction } d  \label{eq:tree_stab_constraints}\\
  &&x & \geq 0\label{eq:nonne}
\end{alignat}
Equation~\eqref{eq:total} ensures the right number of edges in a tree
solution, and connectivity is given by the cut constraints~\eqref{eq:cut}. 
Just like blossom constraints for matching problems, separation over these
constraints can be done in polynomial time by means of a minimum cut
routine, so this LP can also be solved in strongly polynomial time.


\section{Iterated Rounding}
\label{sec:iterated}

The above LP relaxations provide lower bounds for optimal solutions, 
but no upper bounds in the form of feasible solutions for our stabbing 
problems (except if we solve the integer programs, which may take
exponential time). Therefore, our next objective is to find, in
polynomial time, an integer solution 
that is not too far from an optimal one. 

\begin{figure}[hbtp]
  \centering
  \scalebox{.75}{\epsffile{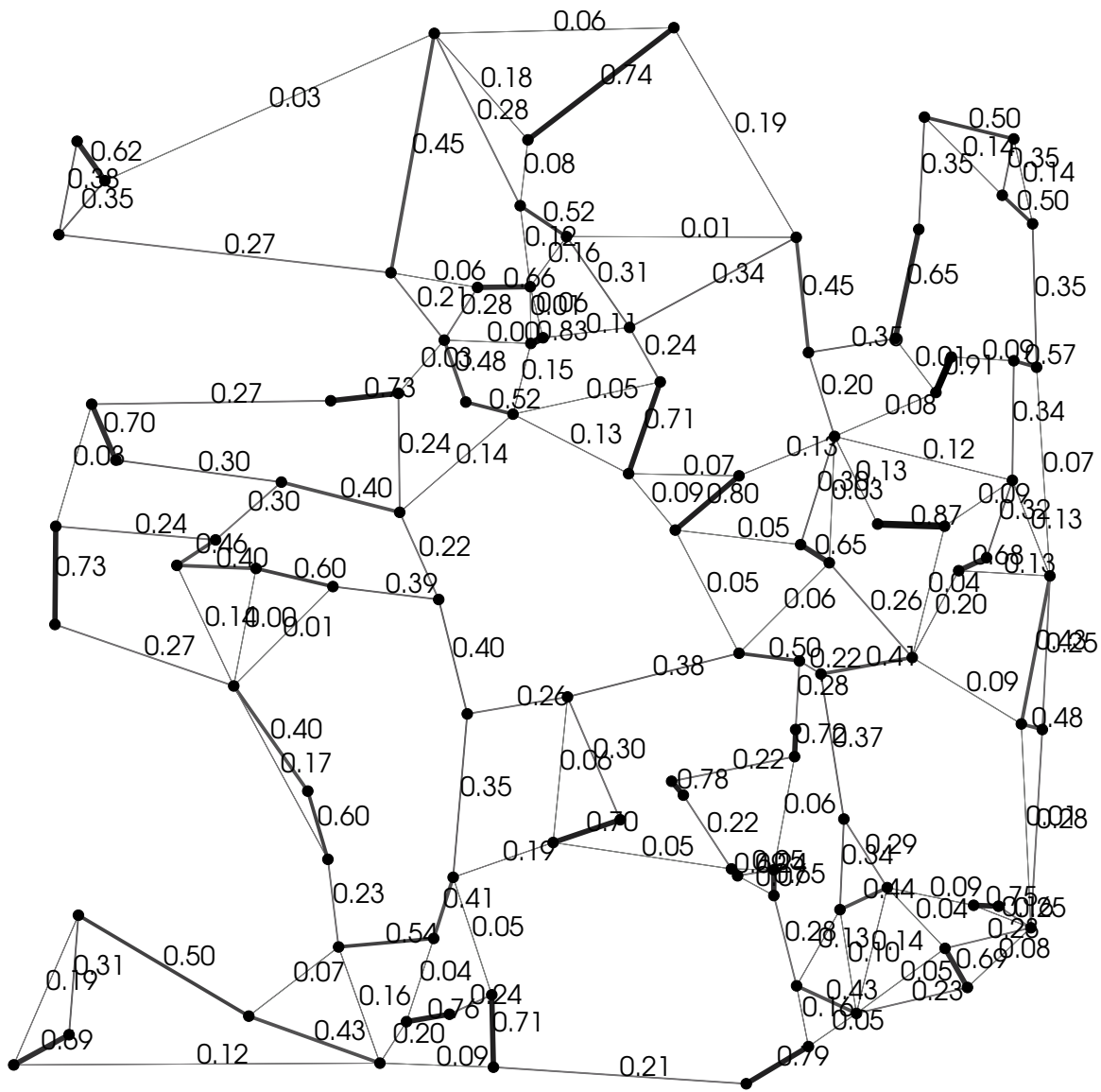}}
  \caption{Finding a fractional matching
of minimum stabbing number: shown is an optimal LP solution
for an 
instance of 100 random vertices; see \cite{flm-csosn-07}
for technical details and results for other 
instances.}
  \label{fig:random100a}
\end{figure}

For this purpose we consider the \emph{support graph} of a (fractional) 
solution $x$: it consists of
all edges $e$ that have a strictly positive value $x_e>0$. 
If we could be assured that in any optimal LP solution,
all edges $e$ in the support graph
solution were $b$-heavy, i.e., $x_e\geq b$ for some constant $b>0$,
then we could consider rounding up all positive edge weights to one
in order to get a $1/b$-approximation. 
Unfortunately, Figure~\ref{fig:random100a} gives an
indication that there may not be such a lower bound.

An alternative to simple LP rounding is offered by the iterated rounding 
technique introduced by Jain~\cite{jain} for generalized Steiner network
problems: 
at each step, it suffices to identify one heavy edge, round it up to one,
fix it, and re-solve the remaining LP. 
The key ingredient is that the modified LP
also has a heavy edge, so the process can be iterated, hence
the name: \emph{iterated rounding}.

A crucial step in Jain's seminal work was the fact
that for so-called generalized Steiner networks,
$1/2$ is a valid lower bound
for the value of the heaviest edge of fractional weight. A sufficient
condition for the validity of this bound that is purely combinatorial
is that all constraints in the LP relaxation are cut constraints, i.e.,
of type $\sum_{e\in\delta(S)}x_e\geq b(S)$.

Unfortunately, stabbing constraints do not fall into the category of 
cut constraints, and it can be seen
from Figure~\ref{fi:third} that Jain's lower bound does not hold for stabbing
problems. This makes it necessary to establish a separate lower  bound
on the value of the heaviest fractional edge. We will do this by exploiting
the underlying geometric nature of our optimization problem, and establish
the fact that there always is an optimal LP solution with planar support
graph. Planarity is proven by shifting weight from a crossing pair of edges
to a non-crossing one; for
matchings (Lemma~\ref{le:planar}) this requires some extra care because of
the blossom inequalities.  The proof for spanning trees
(Lemma~\ref{le:planar.tree}) is almost completely analogous.

\begin{lemma}
\label{le:planar}
For any even set of vertices in the plane, there is a fractional
perfect matching $x$ of minimum stabbing number, such that the 
support graph of $x$ is planar. Such a fractional matching can be
found in polynomial time.
\end{lemma}

\begin{figure}[htb]
 \begin{center}
  \leavevmode
  \centerline{\epsfig{file=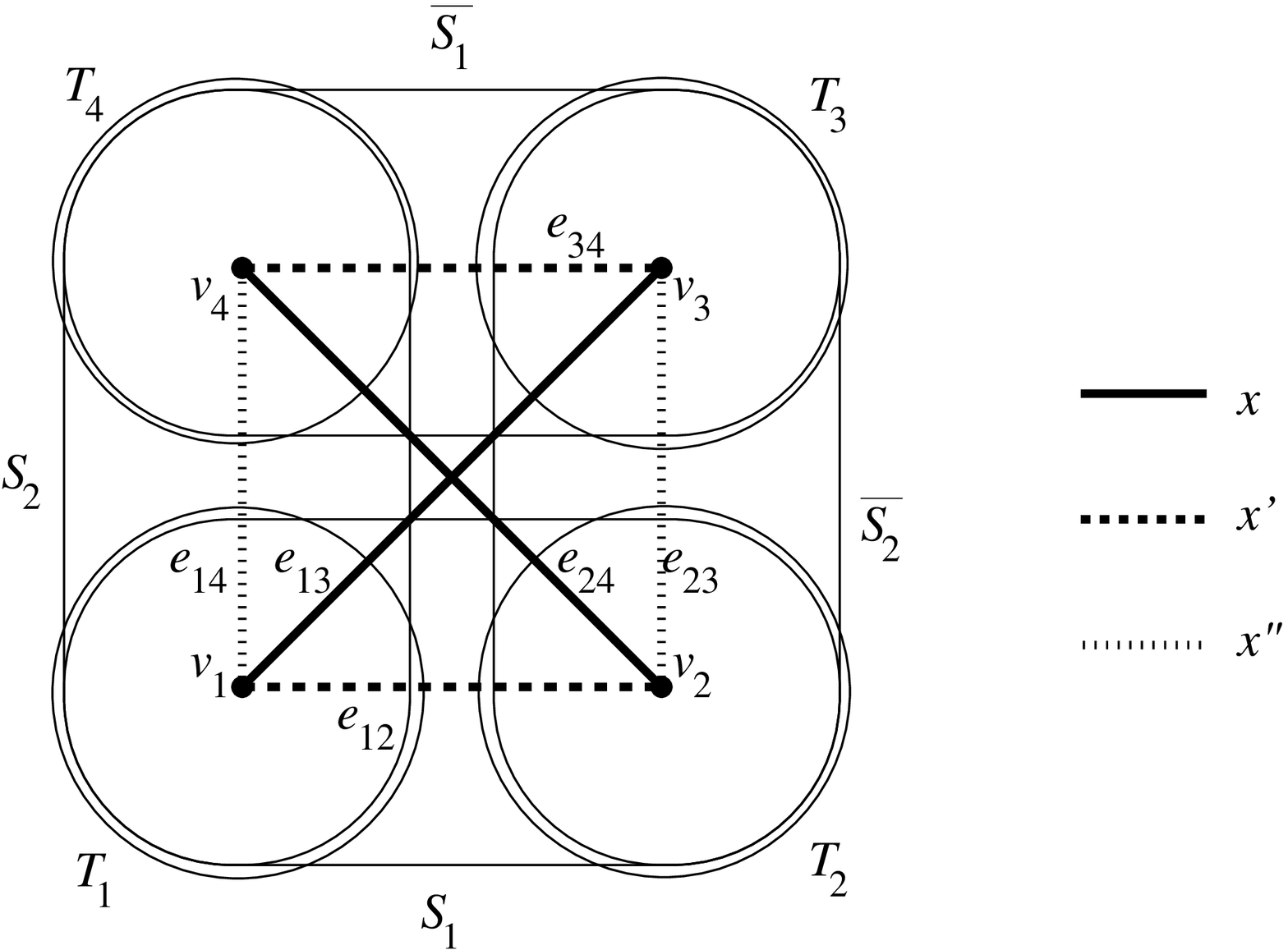, width=.6\textwidth}}
  \caption{Improving a fractional solution with crossing edges
while preserving all blossom inequalities.}
  \label{fig:planar}
 \end{center}
\end{figure}

\proof
The set of all LP solutions is bounded and nonempty, so 
the set of all optimal solutions is nonempty and compact. From this set,
consider a solution $x$ that minimizes the total stabbing number,
i.e., the sum of stabbing numbers over all combinatorially different lines.
We claim that the support
graph of $x$ cannot contain any crossing pair of edges.
Refer to Figure~\ref{fig:planar}.

Suppose there were $e_{13}:=\{v_1,v_3\}$, $e_{24}:=\{v_2,v_4\}$ with 
$x_{e_{13}}> 0$ and $x_{e_{24}}> 0$, such that $e_{13}$ and $e_{24}$ cross. 
We will argue that this implies the existence of an alternative solution
of the same maximum stabbing number, but infinitesimally smaller total stabbing
number, i.e., a contradiction to our assumptions on optimality.

Consider 
$e_{12}:=\{v_1,v_2\}$,
$e_{34}:=\{v_3,v_4\}$,
$e_{14}:=\{v_1,v_4\}$,
$e_{23}:=\{v_2,v_3\}$. Among all LP constraints \eqref{eq:objk}--\eqref{eq:nonneg}, let $s(x)$ be the smallest positive slack (i.e., difference between
left-hand and right-hand side), and choose 
$0<\eps<s(x)$.
As $\sum_{e\in\delta(v_i)}x_e=1$ and $x_{e_{13}}>\eps$, $x_{e_{24}}>\eps$,
we have $x_{e_{12}}<1-\eps$, $x_{e_{34}}<1-\eps$, $x_{e_{14}}<1-\eps$, 
$x_{e_{23}}<1-\eps$. 

Consider two possible alternative solutions arising from
shifting an $\eps$ of weight from $e_{~13}$ and $e_{24}$. 
Let $x'$ and $x''$ be defined by
\begin{equation}
x'_e:=\left\{
\begin{array}{ccc}
x_e-\eps &\mbox{ for }& e\in\{ e_{13}, e_{24}\},\\
x_e+\eps &\mbox{ for }& e\in\{ e_{12}, e_{34}\}\\
x_e &\mbox{ for }& \mbox { all other } e,\\
\end{array}
\right.
\label{eq:x'}
\end{equation}
and 
\begin{equation}
x''_e:=\left\{
\begin{array}{ccc}
x_e-\eps &\mbox{ for }& e\in\{ e_{13}, e_{24}\},\\
x_e+\eps &\mbox{ for }& e\in\{ e_{14}, e_{23}\},\\
x_e &\mbox{ for }& \mbox { all other } e.\\
\end{array}
\right.
\label{eq:x''}
\end{equation}
By convexity, both $x'$ and $x''$
satisfy all stabbing constraints that are valid for $x$.
Furthermore, it is easy to see that
both $x'$ and $x''$ have smaller total stabbing number
than $x$. Thus, it suffices to argue that at least one
of these solutions satisfies all blossom inequalities.

Assume that for each of the two alternative solutions,
a blossom inequality is violated; by our assumption on $\eps$,
this means that there
are two odd sets, $S_1\subset P$ and $S_2\subset P$, such that
$\sum_{e\in\delta(S_1)} x'(e)<1$ and
$\sum_{e\in\delta(S_2)} x''(e)<1$,
while 
$\sum_{e\in\delta(S_1)} x_{e}=1$ and
$\sum_{e\in\delta(S_2)} x_{e}=1$.
Let $s_i:=|\{v_1,v_2,v_3,v_4\}\cap S_i|$. 
It is straightforward to see that for $s_{\,i}\in \{0,1,3,4\}$,
both $x'$ and $x''$ satisfy all blossom inequalities
that are valid for $x$; similarly, the blossom inequalities are satisfied for 
$v_{~1},v_3\in S_1$ and $v_2,v_4\in \overline{S_1}$, 
as $\sum_{e\in\delta(S_1)} x'(e)>\sum_{e\in\delta(S_1)} x(e)$.
Therefore, we only need to consider
$v_{~1},v_2\in S_1$ and $v_3,v_4\in \overline{S_1}$, 
and $v_1,v_4\in S_2$ and $v_2,v_3\in \overline{S_2}$. 
Let 
$T_1:=S_1\cap S_2$,
$T_2:= S_1\cap \overline{S_2}$,
$T_3:= \overline{S_1}\cap \overline{S_2}$
$T_4:= \overline{S_1}\cap S_2$,
and
$E_{ij}:=\{e=\{x,y\}\in E\mid x\in T_i, y\in T_j\}$.
As $|T_1\cup T_2|$,
$|T_1\cup T_4|$,
$|T_2\cup T_3|$,
and $|T_3\cup T_4|$ are all odd, we may assume without loss
of generality that
$|T_1|$ and $|T_3|$ are odd, and 
$|T_2|$ and $|T_4|$ are even.

Then we have 
\begin{eqnarray}
\label{eq:est}
2&=&\sum_{e\in\delta(S_1)} x_{e}+\sum_{e\in\delta(S_2)} x_{e}\\
& = & \sum_{e\in\delta(T_1)} x_{e} + \sum_{e\in\delta(T_3)} x_{e}
+ 2 \sum_{e\in E_{24}} x_{e}\\
&\geq&
\sum_{e\in\delta(T_1)} x_{e} + \sum_{e\in\delta(T_3)} x_{e}
+ 2 x_{e_{24}}\\
&>&
\sum_{e\in\delta(T_1)} x_{e} + \sum_{e\in\delta(T_3)} x_{e}.
\label{eq:west}
\end{eqnarray}
This implies that 
$\min\{\sum_{e\in\delta(T_1)} x_{e},\sum_{e\in\delta(T_3)} x_{e}\}<1$,
contradicting the assumption that $x$ satisfies all cut inequalities.
\qed

The above lemma establishes the existence of optimal non-crossing
matchings; when actually trying to find a non-crossing matching, it suffices to
consider an additional term in the objective
function of our LP that refers to the total length of the edges. 
This ensures that the total length of matching edges
is minimized, avoiding crossings in the first place.  
For more details, see our experimental paper \cite{flm-csosn-07}.

For the case of spanning trees, we can use a similar approach,
based on a similar lemma.

\begin{lemma}
\label{le:planar.tree}
For any set of vertices in the plane, there is a fractional
spanning tree $x$ of minimum stabbing number, such that the 
support graph of $x$ is planar. Such a fractional spanning tree
can be found in polynomial time.
\end{lemma}

\proof
We proceed completely analogous to the proof of the previous lemma
to deduce that if there was a fractional solution $x$ of
minimum total stabbing number among all solutions with
optimal stabbing number, such that the support graph of $x$ 
has two crossing edges, then we could shift an infinitesimal
amount of weight from the crossing edges to two non-crossing ones,
such that the stabbing number remains the same, but the total 
stabbing number decreases. 

The technical steps of the argument
are virtually identical: consider two crossing edges
$e_{13}$ and $e_{24}$ as shown in Figure~\ref{fig:planar}, and
the alternative solutions $x'_e$ and $x''_e$ as defined in
\eqref{eq:x'} and \eqref{eq:x''}. For arguing that at least one of these
solutions is feasible, we do not have to consider blossom
constraints, but cut constraints of type (\ref{eq:cut}). 

Assume that for each of the two alternative solutions,
a cut constraint is violated; by our assumption on $\eps$,
this means that there
are two sets, $S_1\subset P$ and $S_2\subset P$, such that
$\sum_{e\in\delta(S_1)} x'(e)<1$ and
$\sum_{e\in\delta(S_2)} x''(e)<1$,
while 
$\sum_{e\in\delta(S_1)} x_{e}=1$ and
$\sum_{e\in\delta(S_2)} x_{e}=1$.
Let $s_i:=|\{v_1,v_2,v_3,v_4\}\cap S_i|$. 
It is straightforward to see that for $s_{\,i}\in \{0,1,3,4\}$,
both $x'$ and $x''$ satisfy all cut inequalities
that are valid for $x$; similarly, the cut inequalities are satisfied for 
$v_{~1},v_3\in S_1$ and $v_2,v_4\in \overline{S_1}$, 
as $\sum_{e\in\delta(S_1)} x'(e)>\sum_{e\in\delta(S_1)} x(e)$.
Therefore, we only need to consider
$v_{~1},v_2\in S_1$ and $v_3,v_4\in \overline{S_1}$, 
and $v_1,v_4\in S_2$ and $v_2,v_3\in \overline{S_2}$. 
Let 
$T_1:=S_1\cap S_2$,
$T_2:= S_1\cap \overline{S_2}$,
$T_3:= \overline{S_1}\cap \overline{S_2}$
$T_4:= \overline{S_1}\cap S_2$,
and
$E_{ij}:=\{e=\{x,y\}\in E\mid x\in T_i, y\in T_j\}$.
Using a sequence of estimates that is completely identical to 
(\ref{eq:est})--(\ref{eq:west}), we get
$2>\sum_{e\in\delta(T_1)} x_{e} + \sum_{e\in\delta(T_3)} x_{e},$
again implying that 
$\min\{\sum_{e\in\delta(T_1)} x_{e},\sum_{e\in\delta(T_3)} x_{e}\}<1$,
contradicting the assumption that $x$ satisfies all blossom inequalities.
\qed

\begin{theorem}
\label{th:heavy}
For any even set of vertices in the plane, there is a fractional
perfect matching $x$ of minimum stabbing number that has
an edge of weight at least $1/5$.
For any set of vertices in the plane, there is a fractional
spanning tree $x$ of minimum stabbing number that has
an edge of weight more than $1/3$.
\end{theorem}

\proof
For both problems, consider a fractional vertex with a planar support graph.
To see the claim for matchings, note that 
there must be a vertex with degree at most five; as the total
weight for each vertex is 1, the claim follows.
To see the claim for spanning trees, note that 
the total edge weight is $n-1$, and the number of edges
is at most $3n-6$, implying that the average weight is larger
than $1/3$.
\qed

Theorem~\ref{th:heavy} provides the basic
ingredient for an iterated rounding algorithm:
at each iteration, fix the
weight of an edge of maximum fractional weight to one, and re-solve the
linear program. In each iteration, the number of edges with fractional
weight is reduced, so we get an overall polynomial-time algorithm for
finding an integral solution. 

Unfortunately, Jain's original proof only guarantees a constant-factor
approximation for objective functions that arise as the
(weighted) sum of the edge variables.
However, the situation is different for our objective function
which is a \emph{maximum} over certain sums of edge variables,
so an additional argument is needed for establishing a constant-factor
guarantee. We
are hopeful that this argument can be completed some time in the
future \cite{jain.pers}. As we demonstrate in our experimental paper
\cite{flm-csosn-07},
the practical performance seems to be even
better than the theoretically possible guarantees of 5 and 3.


\old{
\section{Computational Results}
\label{sec:comp}

\subsection{Experimental Setup}

We compiled a test suite of various instances on which we evaluated
our linear/integer program \eqref{eq:objk}--\eqref{eq:integ} and the
iterated rounding technique for \Ms{2}.  The suite includes ten
instances with up to 442 vertices from the TSPLIB~\cite{sep:Reinelt:91}
(last vertex removed from odd cardinality instances; therefore the
results reported here are more meaningful than those
in~\cite{flm-msnmstt-04}); the C-class (``clustered'') of Solomon's
instances of the vehicle routing problem \cite{solomon} with 100
vertices each; 25 regular grids with 20 to 360 vertices, based on grids of
size $5\times5$ up to $20\times20$ in which 20\% of the vertices are
removed (chosen uniformly at random); and a set of instances with up
to 100 random vertices in the plane.  Even though we experimented with
available separation routines for blossom inequalities these are not
included in the LPs on which we report below, as solution quality is
already excellent.

Tables~\ref{tab:results} and~\ref{tab:results2} display our results on
a 2.8 GHz Pentium~4 Linux PC with 1 GB main memory, using the commercial
LP/IP solver CPLEX 9.1. For each instance we list its name, which
indicates the number of vertices; this number is reduced by one for odd names to 
allow a perfect matching. Also listed are the optimal
objective function values for the linear (LPopt) and the integer
(IPopt) program, together with the respective CPU time in
seconds. The last column displays the approximate stabbing number
obtained from iterated rounding. For solving the integer programs we
set a time limit of four CPU hours which was exceeded for some large
instances. This is indicated by brackets around the corresponding
value. In that case we report the CPU seconds it takes to obtain the
listed best known solution.  To provide some intuition what
fractional matchings of minimum (fractional) stabbing number look like,
we show several of them in
Figures~\ref{fig:c_1_4_1}--\ref{fig:pcb442}. The edge weight is
proportional to the thickness of edges in the drawing.

\begin{table}[htbp]
  \centering
  \begin{tabular}{lrrrrr}
\hline
Instance&LP opt & LP CPU & IP opt & IP CPU  & iterated\\
\hline
ulysses22&  1.992   &0.00      &  2        &0.01       & 2    \\
berlin52&   2.815   &0.02      &  4        &0.90       & 5    \\
lin105  &   5.500   &0.15      &  6        &80.57      & 8    \\
bier127 &    4.297  &0.34      &  (6)      &(3.90)     & 7    \\
u159    &   15.000  &0.15      &  15        & 2.37     & 15   \\
ts225   &   13.700  &0.35      &  (15)      &(122.28)  & 16   \\
tsp225  &   11.500  &0.32      &  12       &7.66       & 12   \\
a280    &   10.500  &4.26      &  (12)     &(284.80)   & 12   \\
lin318  &  8.113    &12.65     &  (10)     &(6825.48)  & 11   \\
pcb442  &  16.500   &20.71     &  17       &3289.41    & 18   \\
\hline
c101    &      7.000&0.03  &  7&0.54    &     8    \\
c102    &      7.000&0.03  &  7&0.54    &     8    \\
c103    &      7.000&0.03  &  7&0.53    &     8    \\
c104    &      7.000&0.03  &  7&0.53    &     8    \\
c105    &      7.000&0.03  &  7&0.53    &     8    \\
c106    &      7.000&0.04  &  7&0.54    &     8    \\
c107    &      7.000&0.03  &  7&0.54    &     8    \\
c108    &      7.000&0.03  &  7&0.53    &     8    \\
c201    &      6.000&0.03  &  6&2.37    &     7    \\
c202    &      6.000&0.03  &  6&2.37    &     7    \\
c203    &      6.000&0.03  &  6&2.36    &     7    \\
c204    &      6.000&0.03  &  6&2.37    &     7    \\
c205    &      6.000&0.04  &  6&2.35    &     7    \\
c206    &      6.000&0.04  &  6&2.37    &     7    \\
c207    &      6.000&0.04  &  6&2.46    &     7    \\
c208    &      6.000&0.40  &  6&4.18    &     7    \\
\hline                                           
\end{tabular}
  \caption{TSPLIB and clustered instances: comparison of fractional
    and integer optimal stabbing   number \Ms{2}, and the one obtained
    from iterated rounding. Brackets around values indicate an
    exceeded time limit of four CPU hours, and we report the best
    known solution obtained after the time given.}
  \label{tab:results}
\end{table}

\begin{table}[htbp]
  \centering
  \begin{tabular}{lrrrrr}
\hline
Instance&LP opt & LP CPU & IP opt & IP CPU  & iterated\\
\hline
grid5a  &      2.500&0.00  &  3&0.01    &      3    \\
grid5b  &      2.750&0.00  &  3&0.00    &      3    \\
grid5c  &      2.750&0.01  &  3&0.01    &      4    \\
grid5d  &      2.000&0.00  &  3&0.01    &      3    \\
grid5e  &      2.500&0.00  &  3&0.01    &      3    \\          
grid8a  &      5.003&0.01  &  6&0.03    &      6    \\
grid8b  &      5.125&0.01  &  6&0.05    &      6    \\  
grid8c  &      5.000&0.00  &  5&0.05    &      6    \\
grid8d  &      5.429&0.01  &  6&0.04    &      7    \\
grid8e  &      5.403&0.00  &  6&0.21    &      6    \\
grid10a &      4.250&0.01  &  5&0.17    &      6    \\
grid10b &      4.250&0.01  &  5&0.13    &      6    \\
grid10c &      5.250&0.01  &  6&0.19    &      6    \\
grid10d &      4.500&0.02  &  5&1.17   &      6    \\
grid10e &      5.000&0.01  &  5&0.32    &      6    \\
grid15a &      6.000&0.03  &  6&15.92    &      7    \\
grid15b &      7.500&0.03  &  8&1.11    &      8    \\
grid15c &      6.000&0.03  &  (7)&(7.01)    &      7    \\
grid15d &      6.500&0.03  &  7&54.73   &      7    \\
grid15e &      6.750&0.02  &  7&761.33    &      7    \\
grid20a &      9.167&0.98  & (11)&(43.17)  &     12  \\
grid20b &      9.250&0.27  & (11)&(84.97)  &     11  \\
grid20c &      9.500&1.54  & (11)&(33.05)  &     12  \\
grid20d &      9.500&2.24  & (11)&(448.00)  &     12  \\
grid20e &      10.000&2.89 & 11&1169.66  &     12  \\
\hline                                                
random10a  &      1.750&0.00         &  2&0.01    & 2    \\
random10b  &      1.834&0.00          &  2&0.00   & 2    \\
random10c  &      1.750&0.00          &  2&0.01   & 2    \\          
random10d  &      1.700&0.00          &  2&0.01   & 2    \\
random10e  &      1.813&0.00          &  2&0.01   & 2    \\
random50a  &      2.595&0.25          &  3&19.83  & 4    \\
random50b  &      2.628&0.23          &  3&1.91   & 4    \\
random50c  &      2.669&0.23          &  4&30.77  & 4    \\
random50d  &      2.662&0.22          &  4&15.99  & 4    \\
random50e  &      2.790&0.33          &  4&25.54  & 4    \\
random100a &      3.376&5.57          &  (5) & (14.00)  & 6    \\
random100b &      3.406&1.04          &  (5) &(13.81)     &   5        \\
random100c &      3.247&0.99          &  (5)&(16.31)     &  6         \\
random100d &      3.211&0.89          &  (5) &(7.35)     &  6         \\
random100e &      3.233&0.90          &  (5) &(5.84)     &  5         \\
\hline
  \end{tabular}
  \caption{Results for grids and random instances}
  \label{tab:results2}
\end{table}

\begin{figure}[hbtp]
  \centering
  \scalebox{.4}{\epsffile{lp-c_1_4_1.eps}}
  \caption{LP optimal solution for a ``clustered'' instance similar to
    c101, but with 400 vertices}
  \label{fig:c_1_4_1}
\end{figure}

\begin{figure}[hbtp]
  \centering
  \scalebox{.8}{\epsffile{lp-grid20b.eps}}
  \caption{LP optimal solution for grid20b}
  \label{fig:grid20b}
\end{figure}

\begin{figure}[hbtp]
  \centering
  \scalebox{.45}{\epsffile{lp-pcb442.eps}}
  \caption{LP optimal solution for pcb442; `pcb' stands for `printed
    circuit board'}
  \label{fig:pcb442}
\end{figure}

\subsection{Brief (Additional) Observations}

In fractional solutions variables may assume rather arbitrary
fractional and small values; this is also true when blossom
inequalities are added.  The collinearity of vertices in the grid
instances enables us to reduce the number of stabbing constraints,
resulting in significantly reduced computation times. The clustering
of vertices in the vehicle routing instances obviously facilitate the
LP/IP solution process, as was to be expected. However, this
observation is interesting in practice where the data is usually well
structured, as opposed to randomly distributed.

In our experiments, the stabbing number obtained from iterated
rounding is extremely close to the optimum: it is never off by more
than by an \emph{absolute value} of two, i.e., much better than
predicted by our analysis (Lemma~\ref{le:planar}). Computation times
are comparable to solving the linear program because an LP solver will
exploit the fact that linear programs only differ very slightly from
iteration to iteration, and will perform a ``warm start.''
We also experimented with a ``one good shot at once'' approach that is
based on the fact that each fractional matching is the convex
combination of perfect matchings, by finding a maximum weight perfect
matching in the support graph of the LP solution. This tends to give
very good feasible solutions and certainly deserves further
evaluation, both from a computational and from a theoretical vertex of
view.

We also made an experiment (reported in~\cite{flm-msnmstt-04}) to show
that the stabbing constraints seem to completely destroy the
polyhedral structure of the matching polytope. Half of the original
TSPLIB instances we used are infeasible (because they originally had
an odd number of vertices), and this is not detected by the
state-of-the-art CPLEX IP solver within four CPU hours. Iterated
rounding terminates (quickly) in this case with a non-perfect matching
with one vertex unmatched.

Feasible integer solutions of good quality are usually obtained rather
quickly by our integer programs. The time consuming part appears to be
a proof of optimality, but the lower bound increases only slowly in the
branch-and-bound tree. It would be worthwhile to investigate
strengthening the lower bound obtained from the LP relaxation by means
of valid inequalities (``cutting planes'').

}

\section{Notes and Conclusion}
\label{sec:conc}

We have presented the first algorithmic paper on stabbing numbers,
resolving the long-standing open question of complexity, and providing
an approach that appears to be useful in theory and in practice. 
There are a number of interesting open questions.

Our proofs rely on a strong
degeneracy of the vertex set, and it would be interesting to see a
proof for vertices in general position. 

We were not able to extend our \NP-hardness proof
to the case of finding a triangulation of
minimum general crossing number; another interesting issue
is how to convert the \NP-hardness proof for triangulations
of minimum crossing number into an \NP-hardness proof for
minimum stabbing number.

Probably the most intriguing
open question spawned by our work is whether the iterated rounding scheme
suggested by the existence of a heavy
edge in an optimal fractional solution to our linear programs
(Lemmas~\ref{le:planar} and~\ref{le:planar.tree}) does indeed lead to
a constant-factor approximation algorithm. 
Also, the use of the ellipsoid method (at least as a
theoretical argument) is not ``combinatorial,'' which always has to 
be considered a drawback.

Another interesting question is to decide the existence of structures
of small constant stabbing number. As the hardness proof
for deciding the existence of a matching of stabbing number 5 illustrates,
this is still not an easy task. From some solvable special
cases, we only note one; for a proof, see \cite{flm-csosn-07}.

\begin{theorem}
  \label{thm:t-stab_2=2}
  \Ts{2}=2 and \Ms{\all}=2 can be decided in polynomial time.
\end{theorem}

One may also ask for minimizing the {\em average} instead of the
maximum stabbing number, and refer to the average over the whole
continuum of lines intersecting a set of line segments, instead of
just a combinatorial set of representatives.  This, however, amounts
to solving problems of minimum length, with all implications to
hardness and approximation; again, see \cite{flm-csosn-07}
for details.

\begin{theorem}
  \label{thm:average}
  A set of line segments has minimum average
  (axis-parallel, resp.) stabbing number with respect to uniform distribution
  of lines,
  if and only if the overall Euclidean (Manhattan, resp.) length of all line segments is
  minimum. 
\end{theorem}

We remark that a linear program for minimizing the average stabbing
number can be written with a sum in the objective function (instead of
a maximum as we had to model it), allowing us to directly apply our
iterated rounding technique and obtaining the desired approximation
factors of 3 and 5, respectively; see \cite{flm-csosn-07}.

\section*{Acknowledgments}
We thank Joe Mitchell for pushing us to work on this problem,
and repeated discussions that assured further progress.
We also thank Kamal Jain for some discussions on iterated rounding.
We are greatly indebted to two anonymous referees, who put in many hours of
work in going through all the details of this paper, providing
an amazing number of helpful hints, and greatly improving clarity
and accessibility of this paper. 

\bibliographystyle{plain}
\bibliography{stabbing}
\end{document}